\documentclass[twocolumn]{aastex62}
\usepackage[utf8]{inputenc}
\usepackage{amsmath, amsfonts, amssymb, graphics,graphicx, wrapfig, nicefrac}
\usepackage{epsfig}
\usepackage{epstopdf}
\usepackage{multirow}
\usepackage{graphicx}
\usepackage{rotating}
\usepackage{ulem}

\def\h2{H$_2$}
\def\cm2{$\mathrm{cm^{-2}}$}
\newcommand{\kms}{$\mathrm{km\, s^{-1}\, }$}
\newcommand{\jbeam}{$\mathrm{Jy\, beam^{-1}\, }$}

\submitjournal{ApJ}

\shorttitle{S-bearing molecules in Orion KL}
\shortauthors{Luo et al.}

\begin{document}

\title{Sulfur-bearing molecules in Orion KL}

\author{Gan Luo}
\affiliation{National Astronomical Observatories, Chinese Academy of Science,Beijing 100101, People's Republic of China; luogan@nao.cas.cn, siyi.s.feng@gmail.com, dili@nao.cas.cn}
\affiliation{University of Chinese Academy of Sciences, Beijing 100049, People's Republic of China}
\affiliation{CAS Key Laboratory of FAST, NAOC, Chinese Academy of Sciences, Beijing 100101, People's Republic of China}

\author{Siyi Feng}
\affiliation{National Astronomical Observatories, Chinese Academy of Science,Beijing 100101, People's Republic of China; luogan@nao.cas.cn, siyi.s.feng@gmail.com, dili@nao.cas.cn}
\affiliation{CAS Key Laboratory of FAST, NAOC, Chinese Academy of Sciences, Beijing 100101, People's Republic of China}
\affiliation{National Astronomical Observatory of Japan, 2 Chome-21-1 Osawa, Mitaka-shi, Tokyo-to 181-8588, Japan}

\author{Di Li}
\affiliation{National Astronomical Observatories, Chinese Academy of Science,Beijing 100101, People's Republic of China; luogan@nao.cas.cn, siyi.s.feng@gmail.com, dili@nao.cas.cn}
\affiliation{CAS Key Laboratory of FAST, NAOC, Chinese Academy of Sciences, Beijing 100101, People's Republic of China}
\affiliation{NAOC-UKZN Computational Astrophysics Centre, University of KwaZulu-Natal, Durban 4000, South Africa}

\author{Sheng-Li Qin}
\affiliation{Department of Astronomy, Yunnan University, and Key Laboratory of Astroparticle Physics of Yunnan Province, Kunming, 650091, People's Republic of China; slqin@bao.ac.cn}

\author{Yaping Peng}
\affiliation{College of Science, Yunnan Agricultural University, Kunming, 650201, People's Republic of China}

\author{Ningyu Tang}
\affiliation{National Astronomical Observatories, Chinese Academy of Science,Beijing 100101, People's Republic of China; luogan@nao.cas.cn, siyi.s.feng@gmail.com, dili@nao.cas.cn}
\affiliation{CAS Key Laboratory of FAST, NAOC, Chinese Academy of Sciences, Beijing 100101, People's Republic of China}

\author{Zhiyuan Ren}
\affiliation{National Astronomical Observatories, Chinese Academy of Science,Beijing 100101, People's Republic of China; luogan@nao.cas.cn, siyi.s.feng@gmail.com, dili@nao.cas.cn}
\affiliation{CAS Key Laboratory of FAST, NAOC, Chinese Academy of Sciences, Beijing 100101, People's Republic of China}

\author{Hui Shi}
\affiliation{National Astronomical Observatories, Chinese Academy of Science,Beijing 100101, People's Republic of China; luogan@nao.cas.cn, siyi.s.feng@gmail.com, dili@nao.cas.cn}
\affiliation{CAS Key Laboratory of FAST, NAOC, Chinese Academy of Sciences, Beijing 100101, People's Republic of China}

\begin{abstract}
We present an observational study of the sulfur (S)-bearing species towards Orion KL at 1.3\,mm by combining ALMA and IRAM-30\,m single-dish data.
At a linear resolution of $\sim$800\,au and a velocity resolution of 1\,\kms, we have identified 79 molecular lines from 6 S-bearing species. In these S-bearing species, we found a clear dichotomy between carbon-sulfur compounds and carbon-free S-bearing species in various characteristics, e.g., line profiles, spatial morphology, and molecular abundances with respect to $\rm H_2$.
Lines from the carbon-sulfur compounds (i.e., OCS, $^{13}$CS, H$_2$CS) exhibit spatial distributions concentrated around the continuum peaks and extended to the south ridge. The full width at half maximum (FWHM) linewidth of these molecular lines is in the range of 2 $\sim$ 11\,\kms. The molecular abundances of OCS and H$_2$CS decrease slightly from the cold ($\sim$68\,K) to the hot ($\sim$176\,K) regions. In contrast, lines from the carbon-free S-bearing species (i.e., SO$_2$, $^{34}$SO, H$_2$S) are spatially more extended to the northeast of mm4, exhibiting broader FWHM linewidths (15 $\sim$ 26 \kms). The molecular abundances of carbon-free S-bearing species increase by over an order of magnitude as the temperature increase from 50\,K to 100\,K.
In particular, $\mathrm{^{34}SO/^{34}SO_2}$ and $\mathrm{OCS/SO_2}$ are enhanced from the warmer regions ($>$100\,K) to the colder regions ($\sim$50\,K). Such enhancements are consistent with the transformation of SO$_2$ at warmer regions and the influence of shocks.
\end{abstract}


\section{Introduction} \label{sec:intro}
Sulfur (S)-bearing molecules (e.g., H$_2$S, SO, SO$_2$, CS, OCS, and H$_2$CS) have been detected in various star-forming environments, e.g., infrared dark clouds \citep{Turner1973,Ragan2006,Vasyunina2011}, hot molecular cores \citep{Blake1987,Charnley1997,Tercero2010,Feng2015} and shocked regions associated with protostellar objects \citep{Wakelam2005,Podio2015,Holdship2016,Girart2017}. Given that the relative abundance ratios of S-bearing species are highly sensitive to gas temperature and density\citep{Viti2004,Wakelam2011}, they have been used in previous studies to understand the physical environment of several molecular clouds \citep[e.g.,][]{Charnley1997,Pineau1993,Bachiller1997,Hatchell1998, Van2003,Wakelam2011,Esplugues2014,Feng2015}. However, the feasibility of using these species to precisely diagnose the evolutionary stage of a particular star-forming region is still questionable. The main reason is that the main sulfur carriers on the grain mantle are still uncertain. It has long been proposed that H$_2$S \citep{Charnley1997} and/or OCS \citep{Hatchell1998,Van2003} are candidate sulfur grain reservoirs, forming SO, $\mathrm{SO_2}$ and other S-bearing molecules in the gas phase \citep{Podio2014,Esplugues2014,Holdship2016}. However, OCS and $\mathrm{SO_2}$ have been detected or tentatively detected in the interstellar ices \citep{Palumbo1995, Boogert1997}, and $\mathrm{H_2S}$ has not yet been detected in the solid phase.

The Orion Kleinmann-Low Nebula (Orion KL) is a good site for investigating the physical and chemical evolution of high-mass star-forming regions. It is the closest high-mass star-forming region (437 $\pm$ 19 pc; \citealt{Hirota2007}), and it  exhibits rich molecular line emission at the (sub)millimeter wavelengths \citep{Blake1987,Tercero2010,Crockett2014,Feng2015,Frayer2015,Pagani2017,Peng2017,Peng2019}. Observationally, this region is composed of four major components, the hot core, the compact ridge, the plateau, and the extended ridge, which are spatially and kinematically different \citep[e.g.,]{Blake1987,Schilke2001,Tercero2010,Crockett2014}.
The hot core is characterized by molecular lines with $\mathrm{{\upsilon}_{lsr} \approx 3-5}$ \kms and $\mathrm{{\Delta}{\upsilon} \approx 5-10}$ \kms \citep{Blake1987,Tercero2010}. The hot core is a hot ($\mathrm{T_{kin} \geqslant 150 K}$) and dense ($\mathrm{\geqslant 10^7 cm^{-3}}$) gas clump, which bridges the evolution between the natal molecular cloud and the inner newly formed star \citep[e.g., Source I,][]{Hirota2017,Baez2018}. Although the central heating source(s) is still under debate \citep{Vicente2002, Wang2010, Goddi2011, Wright2017, Orozco2017}, it evaporates the molecules from the ice mantle to the gas phase, and changes the chemistry on a relatively short timescale ($\sim 10^4 \sim 10^5$ years) \citep{Bernasconi1996}. Previous studies suggested that the Orion hot core is rich in nitrogen-bearing complex molecules \citep{Caselli1993,Peng2017}.
The compact ridge is less dense ($\mathrm{\geqslant 10^6 cm^{-1}}$) and has lower gas temperature (80-140\,K) than the hot core \citep{Blake1987}. The central velocity is $\mathrm{\sim 7-8}$ \kms, and the full width at half maximum (FWHM) linewidth is $\sim$ 3-5\,\kms. Oxygen-bearing molecules are more abundant in the compact ridge than in the hot core \citep{Tercero2018}.
The plateau harbors two outflows, with a low-velocity bipolar flow (LVF, 18\,\kms) along the northeast-southwest direction and a high-velocity outflow (HVF, 30-100\,\kms) along the northwest-southeast direction. A larger linewidth ($\geqslant$ 20-25\,\kms) was detected towards the plateau than that of compact ridge \citep{Plambeck2009,Zapata2011,Zapata2012,Bally2017,Hirota2017}.
The extended ridge is the most quiescent region in Orion KL, with a kinetic temperature of  50-60\,K \citep{Blake1987,Tercero2010}. Molecular line observations in the millimeter band indicated a $\mathrm{{\upsilon}_{lsr}}$ of $\sim$ 9\,\kms and a linewidth of $\sim$ 3-4\,\kms towards this region.

The Atacama Large Millimeter Array (ALMA) is now providing fruitful archival data, with broad spectral coverage of S-bearing lines at high sensitivity and high angular resolution, allowing us to perform a detailed study of S-bearing chemistry towards the nearest high-mass star-forming region, Orion KL.
In this paper, we combine archival ALMA data with IRAM-30\,m observations towards Orion KL at 1.3\,mm in Section \ref{sec:observation}. In Section \ref{sec:result}, we identify the S-bearing lines and study their line profiles towards different substructures at a linear resolution of $\sim$800 au. We discuss the chemical relations of different S-bearing species according to their spatial variations in Section \ref{sec:discussion}. The conclusions are summarized in Section \ref{sec:conclusion}.

\section{Observations}\label{sec:observation}

\subsection{ALMA observations}

\begin{figure}
  \centering
   \includegraphics[width=2.5in,trim=0 0 0 0,clip,angle=-90]{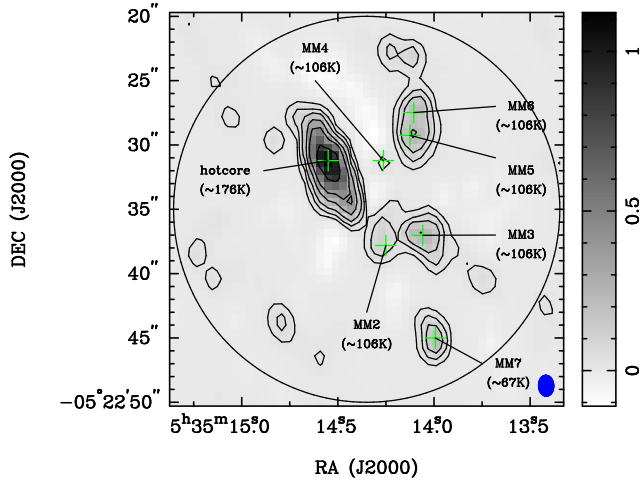}
      \caption{Substructures resolved by ALMA-only continuum observations at 1.3\,mm at a spatial resolution of $1.64'' \times 1.20''$. The black contours show the continuum emission at the 5 $\sigma$, 15 $\sigma$, 30 $\sigma$, 50 $\sigma$, 100 $\sigma$, and 200 $\sigma$ levels with 1 $\sigma = 3.9$ m\jbeam. The green crosses denote the continuum peaks. The black open circle indicates the ALMA primary beam (30$''$). The synthetic beam is indicated in the bottom right by the blue solid ellipse.}
         \label{fig:continuum}
\end{figure}

The ALMA archive data were obtained from ALMA science verification\footnote{https://almascience.nrao.edu/alma-data/science-verification} (SV) data in band 6 (project ID 2011.0.00009.SV). The observations were performed on Jan.~20, 2012, with baselines ranging from 17 to 265 m. The phase-tracking center was at R. A. = $05^h 35^m 14^s.35$ and Dec. = $-05^\circ 22'35''.0$. The ALMA spectral coverage is from 213.719\,GHz to 246.619\,GHz, with a spectral resolution of 0.488\,MHz (corresponding to $\sim$ 0.7\,\kms at 230\,GHz). Bandpass and flux calibration were performed with Callisto. Quasar J0607-085 was observed for phase calibration. Continuum subtraction was performed, and the images were reduced with the MIRIAD \citep{miriad} software. The images were deconvolved with natural weighting using the Clean algorithm. The synthesized beam size is $1.64'' \times 1.20''$ at 230\,GHz. The 1$\sigma$ root mean square (RMS) noise level of the continuum and lines are $\sim$10 m\jbeam and 30 m\jbeam per channel, respectively. Figure \ref{fig:continuum} shows the 1.3\,mm continuum map from the ALMA-only data. Adopting the same nomenclature as that given by \citet{Wu2012}, we label the brightest condensation as the hot core and the resolved condensations as mm2-mm7.

\subsection{Single-dish observation with IRAM}
The ALMA-only observations filtered out $\sim$30-50\% of the extended emission in general compared to the 1.3\,mm line emission detected with the IRAM-30\,m single-dish observations (see the observational details in \citealt{Feng2015}). Therefore, we convert the IRAM-30\,m data into the MIRIAD data format and use the following procedure to combine the ALMA and the IRAM-30\,m data.

\begin{figure}
  \centering
   \includegraphics[width=2.5in,trim=0 0 0 0,clip,angle=-90]{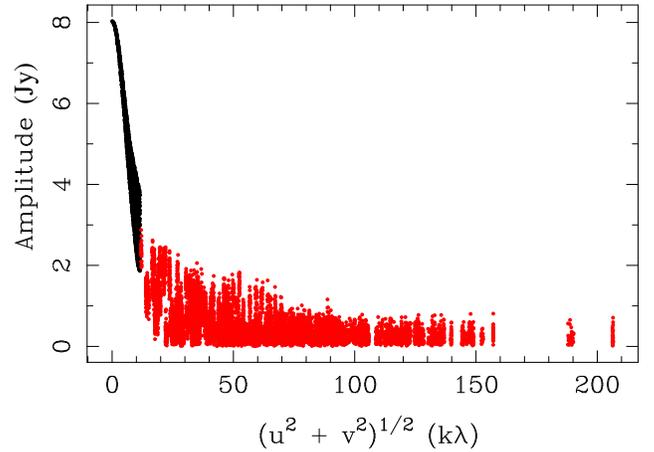}
      \caption{Amplitude as a function of the projected baseline. ALMA data are shown in red and IRAM-30\,m data in black.}
         \label{fig:amplitude}
\end{figure}

\begin{figure}
  \centering
  \includegraphics[width=1.0\linewidth,trim=0 0 0 0,clip,angle=0]{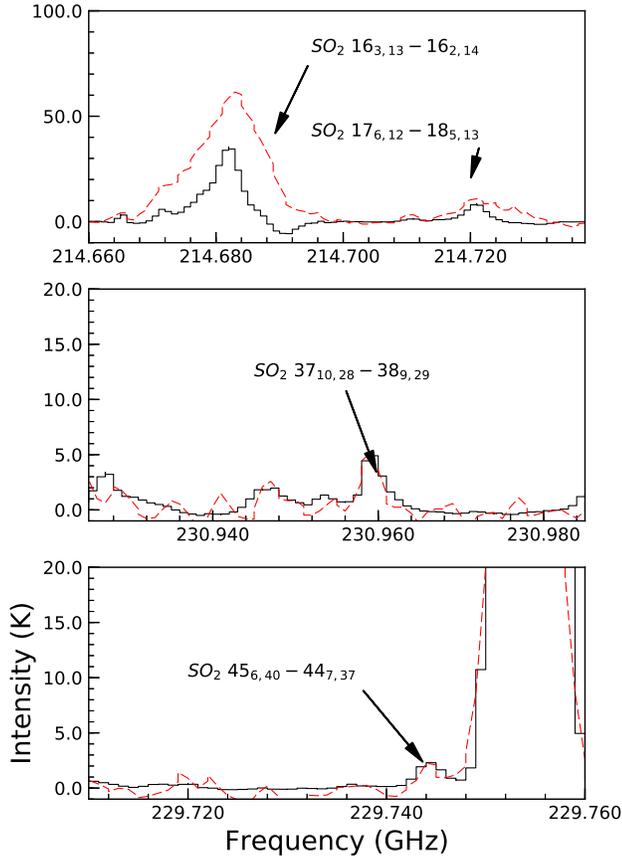}
        \caption{Sample spectra extracted from the peak position towards the hot core. The ALMA-only data are plotted in black and show missing flux in the shape of artificial absorption, and the ALMA-30\,m combined data are plotted in red. The SO$_2$ lines are identified using XCLASS, and the quantum numbers of the unblended transitions are labeled.}
         \label{fig:alma_com}
\end{figure}

First, with the task UVMODEL, model visibility data from the single-dish spectral line are generated in the UV plane. The ALMA visibilities (red dot) and the IRAM-30\,m visibilities (black dot) in the amplitude-UV-distance plane are shown in Figure \ref{fig:amplitude} as an example.
Then, using the Clean algorithm with natural weighting, the ALMA and IRAM-30\,m visibilities are combined into deconvolved images containing large-scale emission (see Figure~\ref{fig:alma_com} for an example of the complementation of missing flux).
The combined datacube has a synthesized beam of $1.86'' \times 1.53''$ (P.A.= $-19^\circ$) in the upper sideband and $2.03'' \times 1.76''$ (P.A.= $-12.6^\circ$) in the lower sideband. The RMS noise level of spectral lines is $\sim$60 m\jbeam per \kms.

\section{Results}\label{sec:result}

\subsection{Line identification}\label{sec:line identification}

The broad bandwidth of the combined dataset covers multiple transitions from a large number of S-bearing species, allowing us to perform an unbiased S-bearing study by excluding excitation effects.  Adopting the eXtended CASA Line Analysis Software Suite (XCLASS, \citealt{xclass}), we are able to identify all lines of a particular species in our data simultaneously, based on molecular databases such as the Cologne Database for Molecular Spectroscopy (CDMS, \citep{Muller2005}) or the database from from the Jet Propulsion Laboratory (JPL, \citep{Pickett1998}).

The spectra extracted from the peak position towards each substructure are shown in Figure \ref{fig:lines}. The hot core contains the largest number of line detections, showing more intensive emission than the remainder of Orion KL. Therefore, using XCLASS, we identified 74 lines from 20 S-bearing isotopologues towards the hot core, including 13 lines of $\mathrm{SO_2}$, 14 lines of $\mathrm{^{34}SO_2}$, 8 lines of $\mathrm{^{33}SO_2}$, 9 lines of $\mathrm{OS^{17}O}$, 9 lines of $\mathrm{OS^{18}O}$, 4 lines of SO, 1 line of $\mathrm{^{34}SO}$, 1 line of $\mathrm{^{33}SO}$, 1 line of $\mathrm{S^{18}O}$, 1 line of H$_2$S, 1 line of $\mathrm{H_2^{34}S}$, 1 line of $\mathrm{H_2^{33}S}$, 2 lines of OCS, 1 line of $\mathrm{OC^{33}S}$, 2 lines of $\mathrm{O^{13}CS}$, 1 line of $\mathrm{^{18}OCS}$, 1 line of $^{13}$CS, 1 line of H$_2$CS, 2 lines of $\mathrm{H_2C^{34}S}$, and 1 line of $\mathrm{H_2C^{33}S}$.

\subsection{Synthetic spectrum fitting}\label{sec:fitting}
For each species, assuming that all lines of that species are under local thermodynamic equilibrium (LTE), we can use the Modeling and Analysis Generic Interface for eXternal numerical codes (MAGIX; \cite{magix}) package to perform the fitting process. Therein, a synthetic spectrum is modeled from an isothermal object in one dimension by taking the optical depth, line blending, source size, velocity, and linewidth into account.

Using MAGIX, we fit the synthetic spectra of all the 20 isotopologues towards each substructure. The input parameters of a particular molecule in MAGIX include the source size (in arcsecond), the rotational temperature T$\mathrm{_{rot}}$ (K), the molecular total column density N$\mathrm{_{tot}}$ (cm$^{-2}$), the FWHM  linewidth $\Delta\upsilon$ (\kms), and the central velocity $\upsilon\mathrm{_{lsr}}$ (\kms). These input parameters are assumed to be the same for different transitions as initial guesses. By minimizing ${\chi}^2$ in the given parameter space, MAGIX yields optimized the results as output. For the isotopologues of the same species, we assume that all lines have the same centroid velocity.
Moreover, we treat cases of line blending in which two or more possible lines contributing more than 5$\%$ of the observed intensity.
An optimal fit is obtained by using three fitting algorithms, genetic, Levenberg-Marquardt, and errorestim-ins. Figure \ref{fig:spw7} shows an example of the synthetic spectra produced by MAGIX fitting to the lines in the frequency range of 215380-217218\,MHz towards individual substructures.

Analyzing the fitting results, we obtain the following results:
\begin{itemize}
\item In general, all the S-bearing species identified in our dataset are well fit, exhibiting small optical depths (Table \ref{tab:spectral_lines}). One exception is SO$_2$ towards the hot core, mm2, mm3, and mm4, where the fits for some low-$J$ lines indicate optical depths greater than 1. These lines may originate from different temperature components (i.e., from both the central protostellar objects and the outer envelope). Photons from the inner hot component will be absorbed by molecules from the colder envelope, which will lead to overfitting of the low-$J$ lines. \citep[see the similar results given in][Appendix B]{Ahmadi2018}. In such cases, we only use the high-$J$ lines for the fitting process. The exclusion of the optically thick (low-$J$) lines will yield more accurate rotational temperatures.

\begin{figure*}
\centering
\includegraphics[width=1.0\linewidth,trim=0 0 0 0,clip]{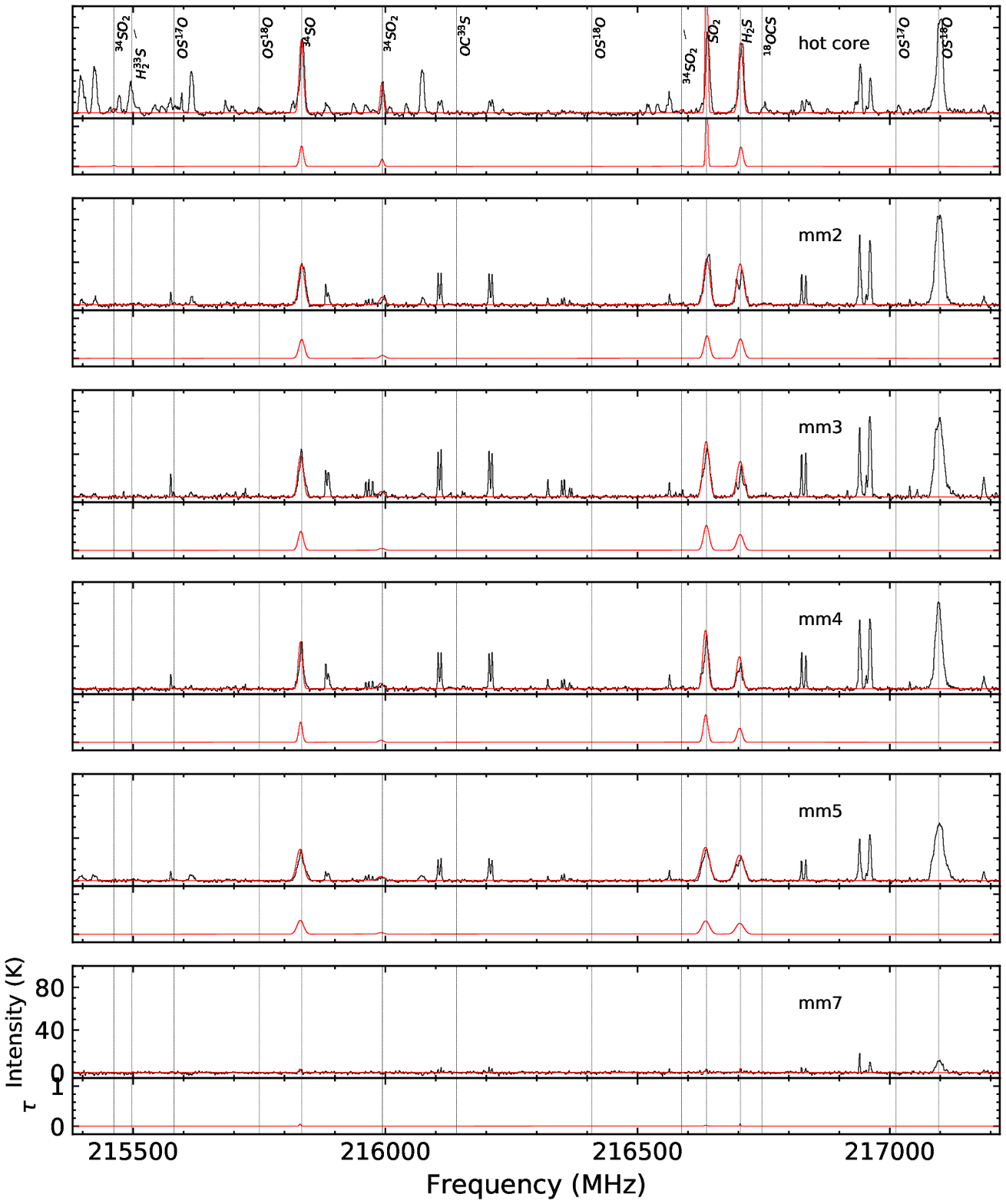}
\caption{An example of the XCLASS fitting results in the frequency range from 215380 to 217218 MHz. The six subfigures show the spectra and optical depths of the identified molecules from six positions. For each subfigure, the black curve in the upper panel shows the combined data, and the red curve shows the model spectra obtained with XCLASS. The red curve in the lower panel shows the optical depth of each line.}\label{fig:spw7}
\end{figure*}

\item The LTE condition seems to be a reasonable assumption for the substructures we study here. Given that a successful fitting of a particular species with MAGIX requests at least three confirmed transitions and that $\mathrm{SO_2}$ is the only species for which we detected more than three unblended lines towards all substructures (see the example of its unblended lines towards the hot core in Figure \ref{fig:alma_com}), we can use its fitting results to validate the assumptions in our source environment when using MAGIX. These $\mathrm{SO_2}$  lines in our dataset cover an $\mathrm{E_u}$ range of 19.03-1126.34\,K.
    The rotation temperature $\mathrm{T_{rot}}$ of $\mathrm{SO_2}$ from the fitting results (Table \ref{tab:parameters}) is shown in Figure~\ref{fig:rt}. We found that $\mathrm{SO_2}$ exhibits the highest $\mathrm{T_{rot}}$ towards the hot core region($\sim 176\,K$) and the lowest $\mathrm{T_{rot}}$ towards mm7 region( $\sim 68\,K$). The $\mathrm{T_{rot}}$ of $\mathrm{SO_2}$ is $\rm \sim106$\,K towards mm2, mm3, mm4, and mm5, without significant variations. This result is consistent with \citealt{Feng2015}. At such high temperature, the critical densities of these lines are $\rm 10^4 \sim 10^6\,cm^{-3}$, which is less than the number density of the substructures ($\mathrm{10^7 \sim 10^8\,cm^{-3}}$). For the rest of the S-bearing species (given in Table \ref{tab:spectral_lines}), we assume that their excitation temperatures towards individual substructures are the same as those of $\mathrm{SO_2}$. Thus, LTE condition is also a good approximation for estimating their column densities.

\begin{figure}
\centering
\includegraphics[width=1.0\linewidth,trim=0 0 0 0,clip]{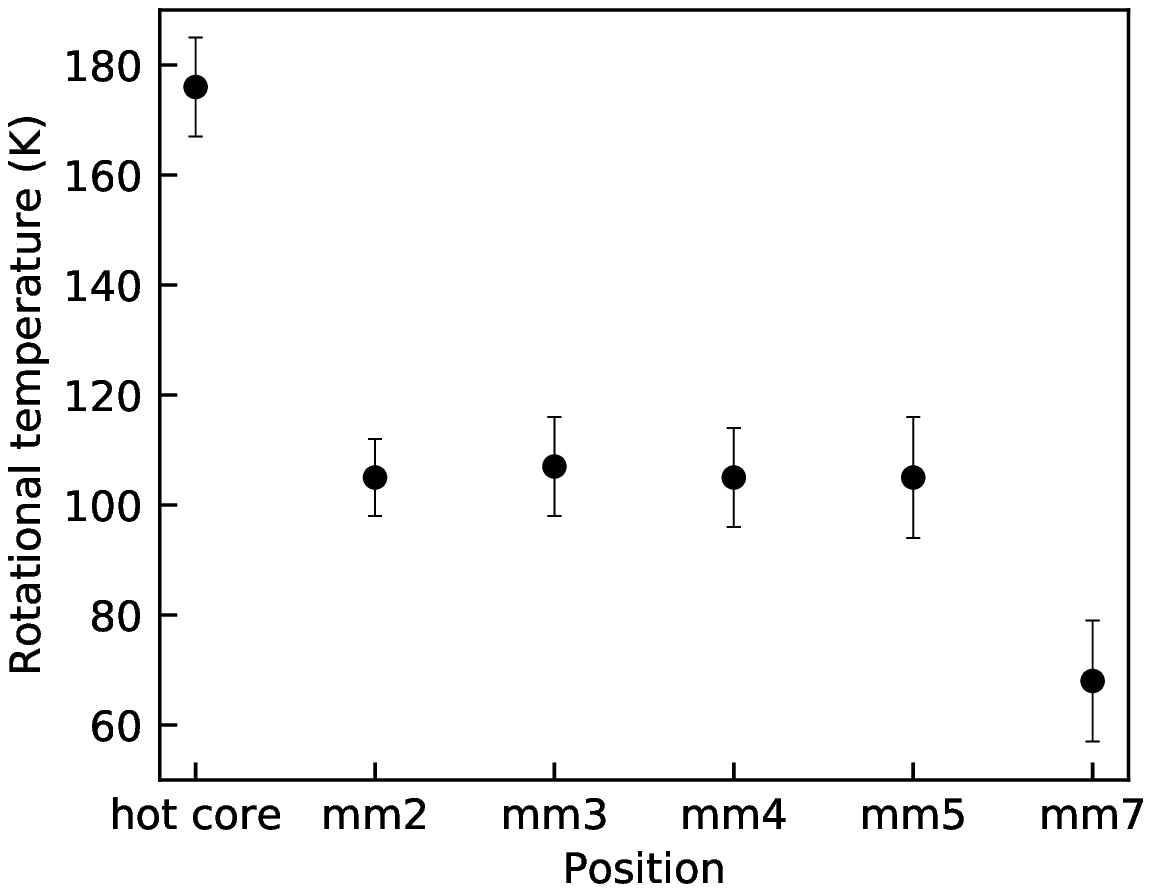}
\caption{The rotational temperatures of $\mathrm{SO_2}$ towards the six substructures derived using MAGIX. The optimal parameters for all S-bearing species are listed in Table \ref{tab:parameters}.}\label{fig:rt}
\end{figure}

\item  The S-bearing lines in our dataset seem to show different line profiles. Figure~\ref{fig:line profile} shows line profiles of representative lines for each species. Figure \ref{fig:linewidth} gives the statistical FWHM linewidth of these species towards individual substructures.
    Most lines from the carbon-free S-bearing species (including $\mathrm{SO_2}$, $\mathrm{^{34}SO_2}$, $\mathrm{^{34}SO}$, and H$_2$S) exhibit a single peak towards the hot core, with a central velocity of $\sim$ 7\,\kms. Those lines have large line wings towards mm2, mm3, mm4, and mm5. Due to overlap of different velocity components along the line-of-sight, the line profiles of H$_2$S show multiple peaks towards mm2, mm3, mm4, and mm5. The FWHM linewidth towards the hot core is 7--14\,\kms and towards mm7 is 3--8 \kms (The FWHM linewidth of $\mathrm{^{34}SO_2}$ is 15 \kms, due to its weak emission), which is significantly narrower than the values for the rest of the substructures (12 $\sim$ 26 \kms).
    Lines from the carbon-sulfur compounds (including OCS, $\mathrm{^{13}CS}$, and $\mathrm{H_2CS}$) exhibit a single velocity component towards each substructure, with a central velocity in the range of $\rm 7 \sim 9\,km\,s^{-1}$. However, they exhibit a narrower FWHM linewidth towards mm2 to mm7 (2$\sim$8 \kms) than that of the hot core (7$\sim$11 \kms).
    Since Orion KL is a complex region with multiple outflows, bringing layers of time-dependent shocks  \citep{Zapata2011,Crockett2014,Bally2017}, the varying linewidths of different S-bearing groups may be the result of components with different chemical ages.

\begin{figure*}
  \centering
  \includegraphics[width=1.0\linewidth,trim=0 0 0 0,clip]{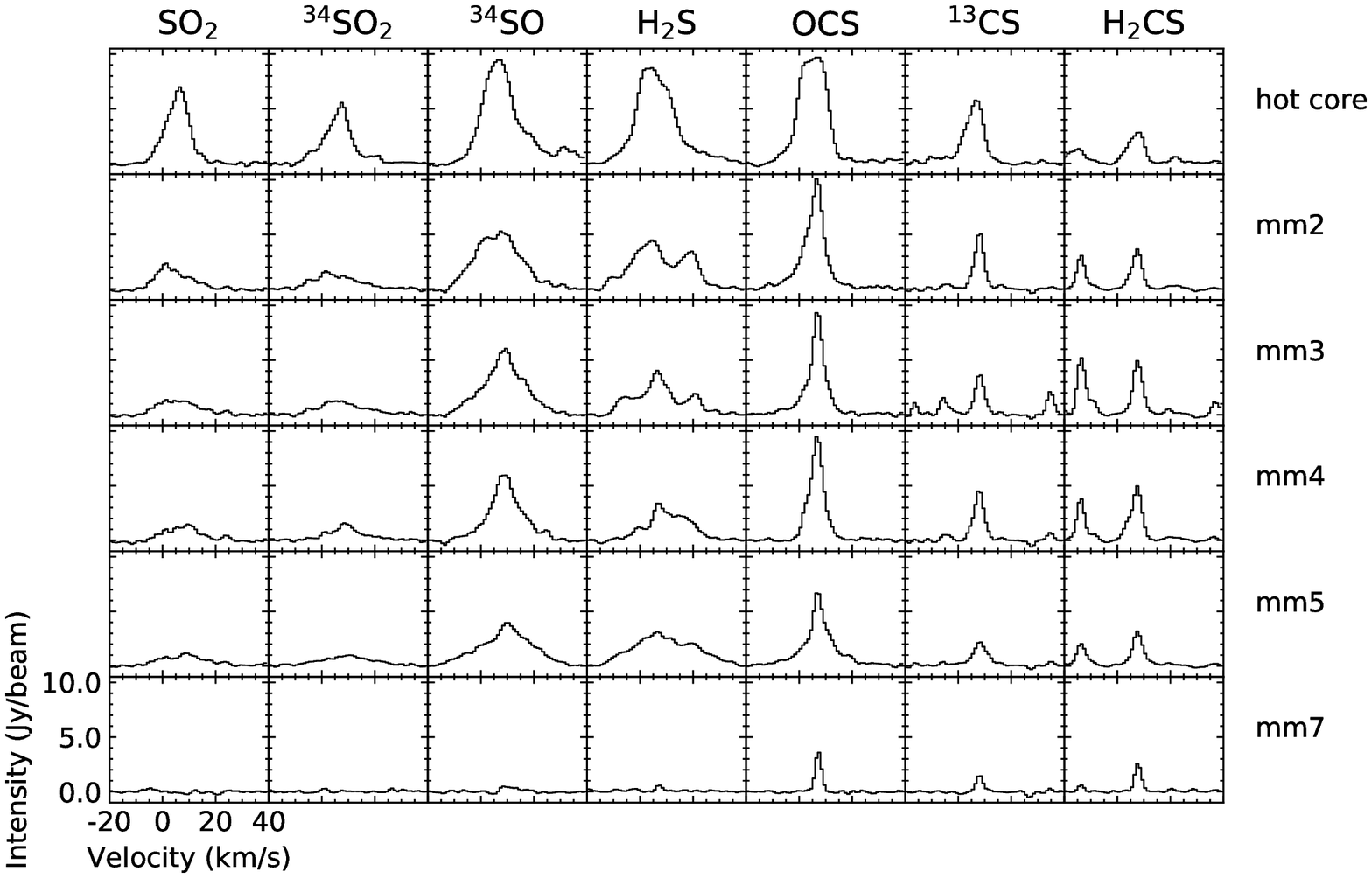}
        \caption{Line profiles of $\mathrm{SO_2}$ ($17_{6,12}$--$18_{5,13}$), $\mathrm{^{34}SO_2}$ ($14_{3,11}$--$14_{2,12}$), $\mathrm{^{34}SO}$ ($6_5$--$5_4$), OCS (18--17), $\mathrm{O^{13}CS}$ (18--17), $\mathrm{^{13}CS}$ (5--4), $\mathrm{H_2S}$ ($2_{2,0}$--$2_{1,1}$), and $\mathrm{H_2CS}$ ($7_{1,7}$--$6{1,6}$) in different regions.}
         \label{fig:line profile}
\end{figure*}

\begin{figure}
  \includegraphics[width=1.0\linewidth,trim=0 0 0 0,clip]{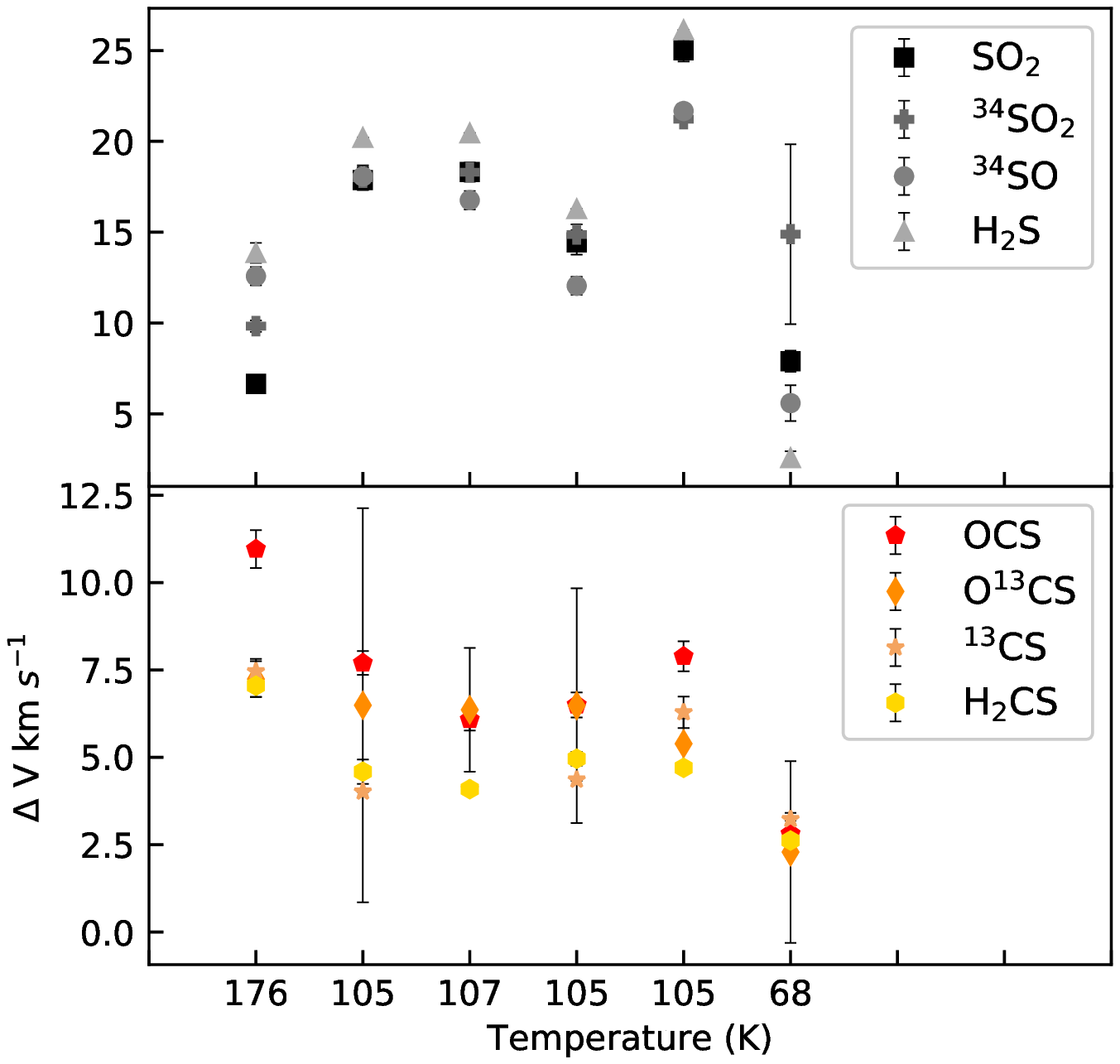}
        \caption{Linewidths of $\mathrm{SO_2}$ ($17_{6,12}$--$18_{5,13}$), $\mathrm{^{34}SO_2}$ ($14_{3,11}$--$14_{2,12}$), $\mathrm{^{34}SO}$ ($6_5$--$5_4$), OCS (18--17), $\mathrm{O^{13}CS}$ (18--17), $\mathrm{^{13}CS}$ (5--4), $\mathrm{H_2S}$ ($2_{2,0}$--$2_{1,1}$), and $\mathrm{H_2CS}$ ($7_{1,7}$--$6{1,6}$) towards six peak positions. The x-axis represents the rotational temperatures towards the hot core to mm7.}
         \label{fig:linewidth}
\end{figure}

\end{itemize}

\subsection{Line spatial distribution}\label{sec:gas distribution}
\begin{figure*}
  \centering
   \includegraphics[width=1.0\linewidth,trim=0 0 0 0,clip,angle=0]{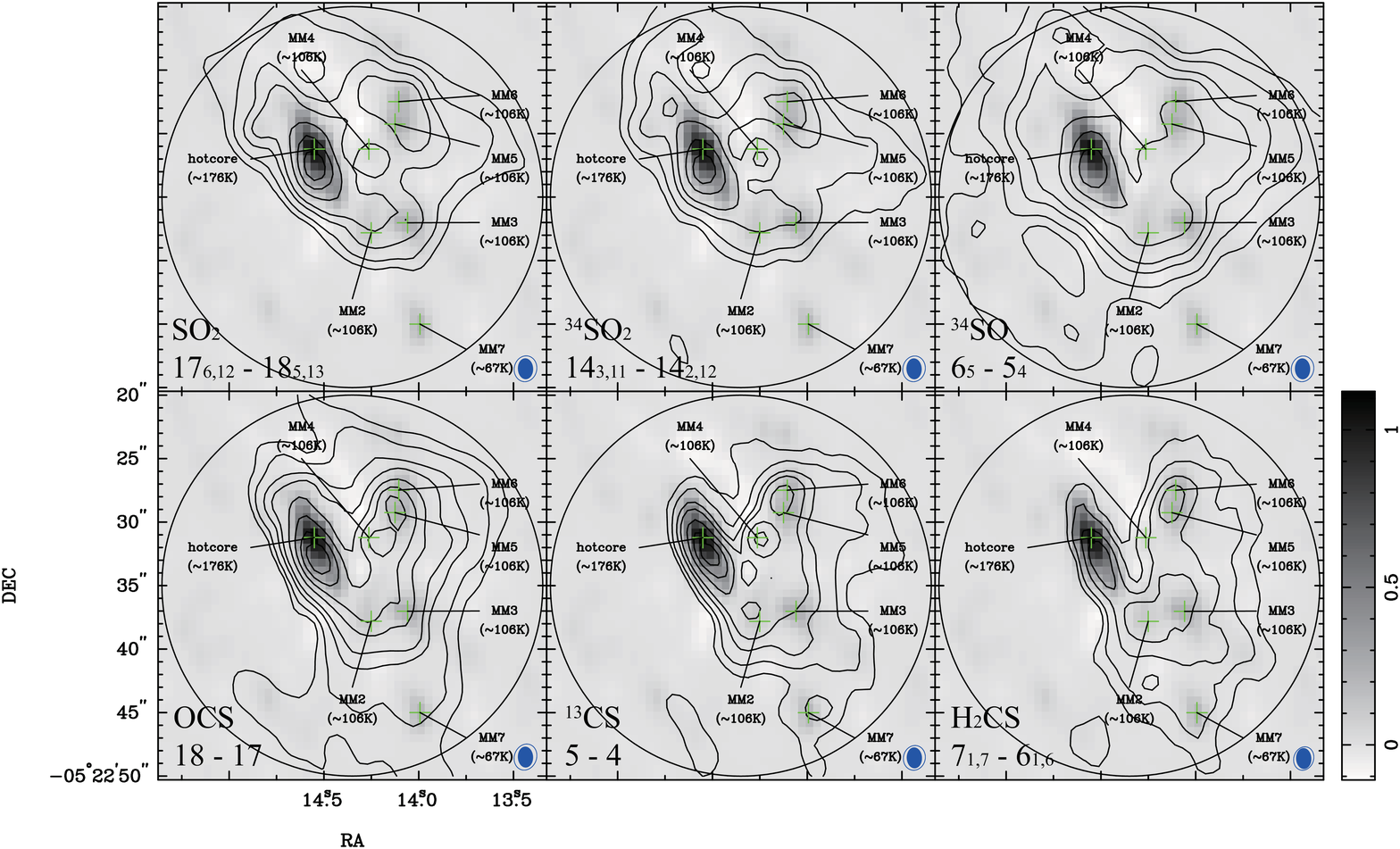}
      \caption{Integrated intensity maps of the representative lines from six species over the velocity range from 0 to 16 \kms. The gray map in the background shows the ALMA-only continuum emission. The peak positions of the continuum emission are marked with green crosses. Black contours show the line emission from the ALMA-30\,m combination, starting from 5 $\sigma$ and increasing by a step of 5 $\sigma$. The transition of each molecule is given in the lower-left corner of each panel.}
         \label{fig:gasmap}
\end{figure*}

Integrating the intensity of the representative lines for each species in the velocity range of 0-16 \kms, we present the spatial distribution maps of six species in Figure \ref{fig:gasmap}. The carbon-free S-bearing species and the carbon-sulfur compounds exhibit different spatial extents.
The extended emissions of the carbon-free species cover the region from northeast mm4 down to the south mm2. The extended emissions of the carbon-sulfur compounds do not cover the northern mm4 but instead shift to the southern mm7, exhibiting a ``heart-shaped" morphology. This result indicates that these two groups of S-bearing species are chemically different, tracing different gas.

\begin{figure*}
  \centering
   \includegraphics[width=4.0in,trim=0 0 0 0,clip,angle=-90]{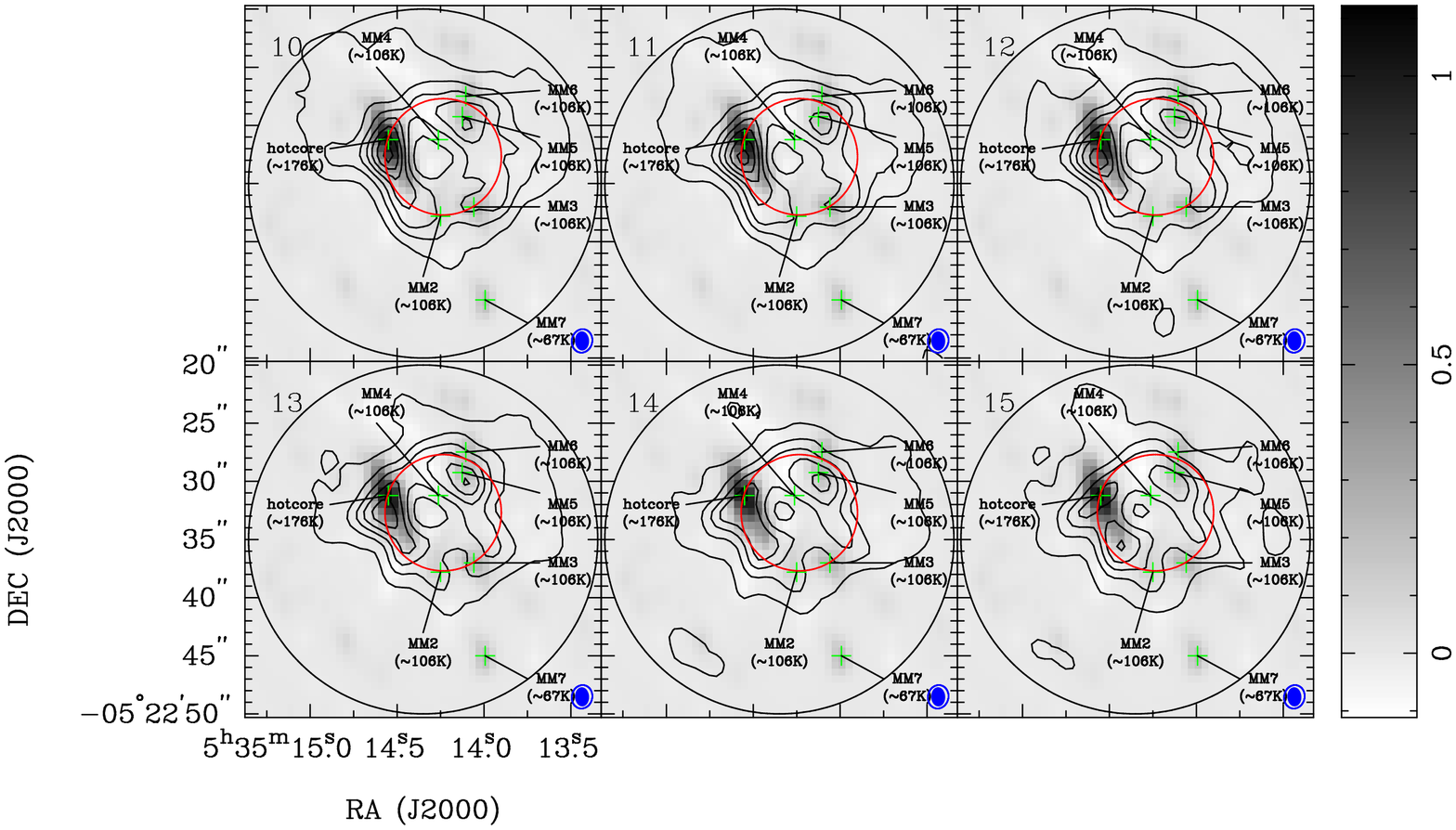}
      \caption{Channel maps of H$_2$S over a velocity range of 10 to 15 \kms. The gray map on the background shows the ALMA-only continuum emission. The peak positions of the continuum emission are marked with green crosses. Black contours show the line emission from the ALMA-30\,m combination, starting from 5\% of the emission peak and increasing by a step of 5\%. The red circle indicates the ring-like structure.\label{fig:H2Schannelmap}}
\end{figure*}

Moreover, when checking the line spatial distribution maps channel by channel, we also note that a ring-like structure appears on the line maps of carbon-free S-bearing species (SO$_2$, $^{34}$SO, H$_2$S) in the velocity range of 10 to 15 \kms (Figure \ref{fig:H2Schannelmap}, \ref{fig:SO2channelmap}, and \ref{fig:34SOchannelmap}). This ring, centered at R. A. = $05^h 35^m 14^s.235$ and Dec. = $-05^\circ 22'32''.7$, has a radius of $\sim$5$''$ in the plane of the sky. The dust emission peaks for the hot core, mm2, mm3, and mm5 are at the edge of the ring, suggesting that the carbon-free S-bearing species may be excited by shocks from the OMC1 explosion 500 years ago \citep{Plambeck2016,Bally2017}. 

\begin{figure*}
  \centering
   \includegraphics[width=1.0\linewidth,trim=0 0 0 0,clip,angle=0]{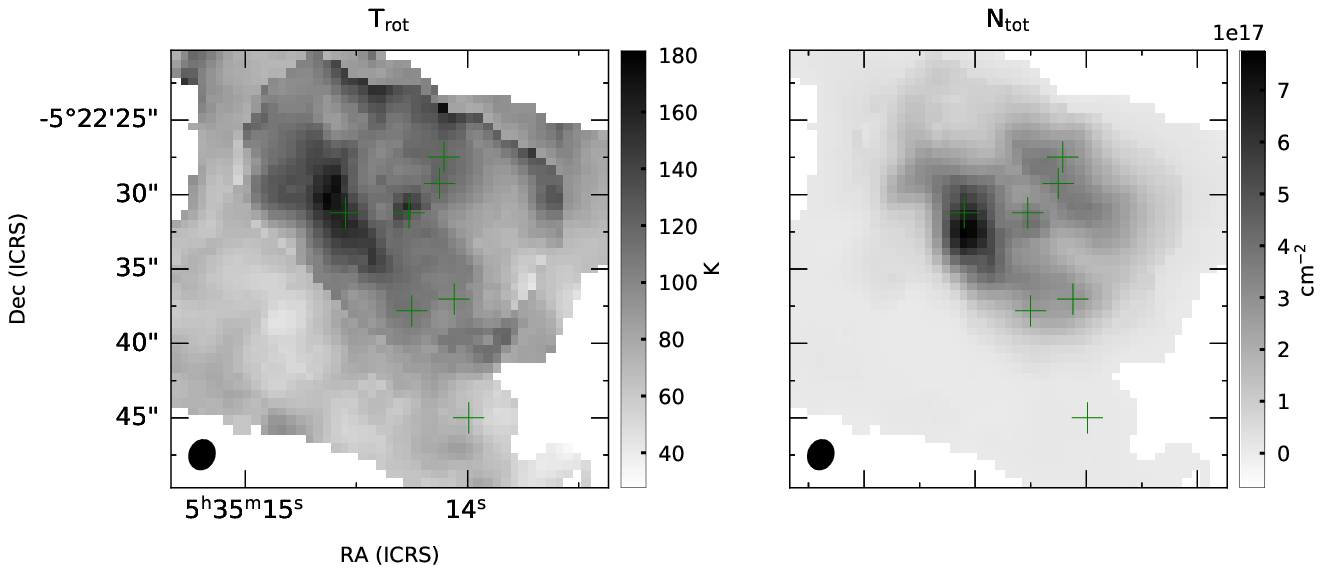}
      \caption{The rotational temperature (left) and column density (right) maps of $\rm SO_2$ obtained using XCLASSMapfit. The molecular lines we used for synthetic fitting are SO$_2$ J = $4_{2,2}$--$3_{1,3}$, $11_{5,7}$--$12_{4,8}$, $16_{1,15}$--$15_{2,14}$, $16_{3,13}$--$16_{2,14}$, $17_{6,12}$--$18_{5,13}$, and $22_{7,15}$--$23_{6,18}$. The continuum peaks are labeled as green crosses. Regions outside the pixel value corresponding to a 3$\sigma$ detection of the most extended integrated intensity of SO$_2$ lines are masked out.}
         \label{fig:so2_map}
\end{figure*}

Using XCLASSMapfit (with the same assumptions and algorithms as MAGIX), we fit the $\mathrm{T_{rot}}$ and column density maps by modeling the synthetic spectra towards all pixels. From the map fittings to SO$_2$ lines (Figure \ref{fig:so2_map}), we note a significant temperature and column density gradient from the hottest (100 $\sim$ 180\,K) and densest (7.8$\times$10$^{17}$ \cm2) center of the hot core, through the warm (100 $\sim$ 120\,K) mm2-mm5, and to the most distant and coolest ($\sim$ 60\,K, 8$\times$10$^{15}$ \cm2) mm7. This gradient may be the result of radiative pumping from being externally heated \citep{Buizer2012,Orozco2017} or shocks \citep{Zapata2011,Wright2017}.

\section{Discussion}\label{sec:discussion}
At a velocity resolution of 1 \kms and a linear resolution of $\sim$800\,au, two groups of S-bearing species exhibit different kinematics and spatial distributions towards the Orion KL complex. These variations indicate that carbon-free S-bearing species and carbon-sulfur compounds may be able to distinguish substructures that have different physicochemical processes in Orion KL.

\subsection{Error budget}\label{sec:iso ratio}
A precise measurement of the molecular column density is key to study the chemical variations in different species. Under the LTE assumption, the column density of $\mathrm{SO_2}$ is obtained by fitting multiple optically thin lines using MAGIX (As stated in Section \ref{sec:fitting}). Column densities of the other S-bearing species are estimated by using the $\rm SO_2$ rotation temperature as the gas temperature. Therefore, to test if these assumptions are reasonable, we derive the relative abundance ratios between the $\rm{^{32}S}$-/$\rm {^{34}S}$-/$\rm {^{33}S}$- isotopologues as well as the $\rm{^{12}C}$-/$\rm{ ^{13}C}$- isotopologues (Table \ref{tab:sulfur ratio}). We compare the values with the canonic $\mathrm{^{32}S/^{34}S}$, $\mathrm{^{34}S/^{33}S}$, $\mathrm{^{12}C/^{13}C}$ isotopic ratios, respectively \citep{Anders1989,Langer1990,Langer1993,Chin1996,Lucas1998,Persson2007,Tercero2010}.

We find that the average relative abundance ratio of $\mathrm{^{32}SO_2/^{34}SO_2}$ towards Orion KL is $20 \pm 4$, which is consistent with previous observations towards Orion KL (e..g, 23 $\pm$ 7 from $\mathrm{^{32}SO_2/^{34}SO_2}$, \citealt{Persson2007}; 20 $\pm$ 6 from $\mathrm{OC^{32}S/OC^{34}S}$, \citealt{Tercero2010}). The result is consistent with the solar value of 23 \citep{Anders1989}, the local diffuse ISM value of 19 $\pm$ 8 from $\mathrm{C^{32}S/C^{34}S}$ from absorption observations \citep{Lucas1998}, and the galactic average value of 24$\pm$5 from $\mathrm{C^{32}S/C^{34}S}$ \citep{Chin1996}.
The average relative abundance ratio of $\mathrm{H_2C^{32}S/H_2C^{34}S}$ is 53$\pm$7 in our study, which is slightly greater than the value of $\mathrm{^{32}S/^{34}S}$ from $\mathrm{^{32}SO_2/^{34}SO_2}$ and a previous study by \citealt{Tercero2010}. The ratio of $\mathrm{H_2C^{32}S/H_2C^{34}S}$ towards the hot core is 19$\pm$8, which is consistent with pervious results. We notice that the linewidth of $\mathrm{H_2C^{34}S}$ is less than 2 \kms (2 channels in the spectra) towards mm2, mm3, mm4, and mm5, which means that the column density of $\mathrm{H_2C^{34}S}$ may be underestimated. Thus, the $\mathrm{^{32}S/^{34}S}$ ratio derived from $\mathrm{^{32}SO_2/^{34}SO_2}$ (20$\pm$4) should be more reasonable.

The average abundance ratio of $\mathrm{^{34}S/^{33}S}$ derived from $\mathrm{^{32}SO/^{34}SO}$ is 6 $\pm$ 1 towards Orion KL, which is consistent with the same molecular pair of \citealt{Esplugues2013} ($\mathrm{6 \pm 3}$), \citealt{Tercero2010} ($\sim$ 5 ), and \citealt{Persson2007} (4.9). Our result is consistent with the solar value of 5.6 \citep{Anders1989} and the galactic average value of 6$\pm$1 from $\mathrm{C^{34}S/C^{33}S}$ \citep{Chin1996}.
Moreover, the derived average $\mathrm{^{12}C/^{13}C}$ ratio (38 $\pm$ 9 from $\mathrm{O^{12}CS/O^{13}CS}$) is consistent with previous work of \citealt{Tercero2010} (45 $\pm$20 from $\mathrm{O^{12}CS/O^{13}CS}$) and \citealt{Persson2007} (57 $\pm$ 14 from $\mathrm{{12}CH_3OH/^{13}CH_3OH}$). \citealt{Langer1990,Langer1993} found a galactic gradient in the $\mathrm{^{12}C/^{13}C}$ isotopic ratio, which increases from 30 in the inner galaxy (5\,kpc) to 70 at 12\,kpc. The $\mathrm{^{12}C/^{13}C}$ ratio they derived from $\mathrm{^{12}C^{18}O/^{13}C^{18}O}$ is 63$\pm$6 towards Orion A, which is greater than our result.

From the above test, we believe that the assumptions of LTE in XCLASS and gas temperature in this work are reasonable. Therefore, the chemical variations between different species can be directly indicated by comparing their relative abundances at the same position.

\subsection{Chemical segregation of the carbon-free S-bearing species and carbon-sulfur compounds}\label{sec:abundance}
In this work, we estimate the relative abundance (column density ratio) of S-bearing species with respect to $\mathrm{H_2}$. Instead of using continuum emission and the gas-to-dust ratio, we use $\mathrm{C^{18}O}$ to estimate the $\mathrm{H_2}$ column density for the following reasons:
(1) The emission lines at 1\,mm are so rich towards Orion KL that it is difficult to define the ``line-free" part for the continuum. Moreover, we do not have the bolometric data to compensate for the missing flux of the ALMA-only continuum observations. Therefore, we do not know how significantly the dust continuum we are currently using is overestimated or underestimated.
(2) The extended emission from the ALMA $\mathrm{C^{18}O}$ (2-1) data is complemented using IRAM-30\,m data, and our MAGIX fitting results indicate that this line is optically thin ($\tau \leq 0.2$). At a kinetic distance of $\rm \sim400$\,pc, the $\mathrm{H_2}$ column density can be estimated by converting from $\mathrm{C^{18}O}$ as [C$^{18}$O/H$_2$] $\sim 2 \times 10^{-7}$ towards Orion KL \citep{Frerking1982,Plume2012,Crockett2014,Giannetti2014}.

\begin{figure*}
  \centering
   \includegraphics[width=1.0\linewidth,trim=0 0 0 0,clip,angle=0]{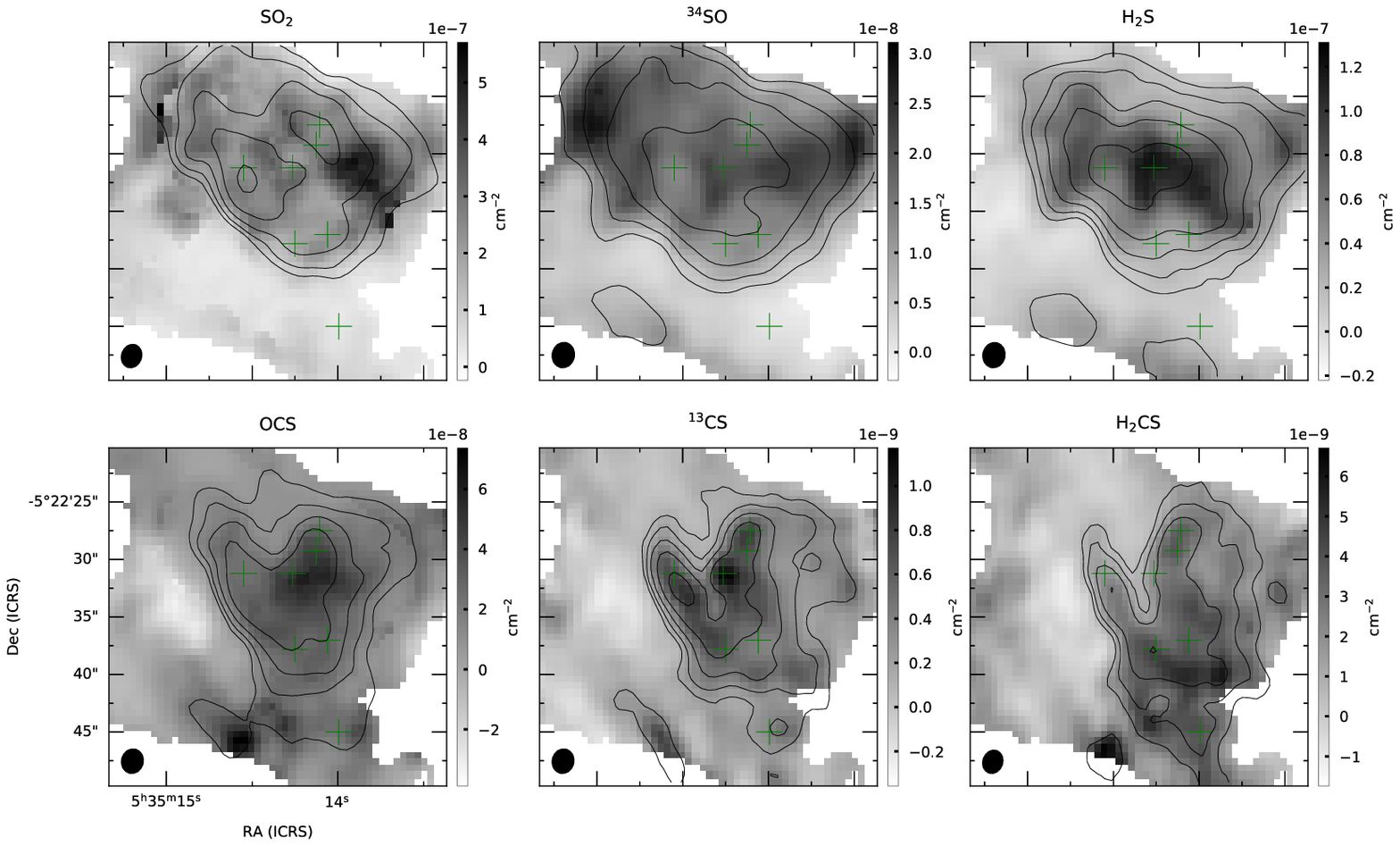}
      \caption{The relative abundance map of $^{34}$SO$_2$, $^{34}$SO, H$_2$S, OCS, $^{13}$CS, and H$_2$CS with respect to $\rm H_2$ (converted from $\rm C^{18}O$ obtained by assuming $\rm C^{18}O/H_2$ ratio of $\rm 2 \times 10^{-7}$). The black contours show the integrated intensity of each molecule in Figure \ref{fig:gasmap}. The contour levels are 1, 2, 4, 8, 16 $\times$ 5$\sigma$. Green crosses label the continuum peaks. Regions outside the pixel value corresponding to a 3$\sigma$ detection of the most extended integrated intensity of SO$_2$ lines are masked out. \label{fig:ab_map}}
\end{figure*}

The column density maps of the S-bearing species are shown in Figure~\ref{fig:col_map}, and their relative abundance ratios with respect to $\rm H_2$ are shown in Figure \ref{fig:ab_map}. In general, the carbon-free S-bearing species exhibit larger relative abundance with respect to $\rm H_2$ to the north, i.e., from the hot core and mm4 to mm5 and mm6, with low abundances towards mm2 and mm7. In contrast, the carbon-sulfur compounds seem to have large abundance not only towards the continuum peaks but also towards mm2 and mm7. Given that the lines we used to derive the molecular column densities are observed simultaneously and cover a large range of $\rm E_U/k$, we believe this chemical segregation is neither an excitation effect nor a sensitivity bias but rather the result of chemical differentiation.

\begin{figure}
  \centering
   \includegraphics[width=1.0\linewidth,trim=0 0 0 0,clip,angle=0]{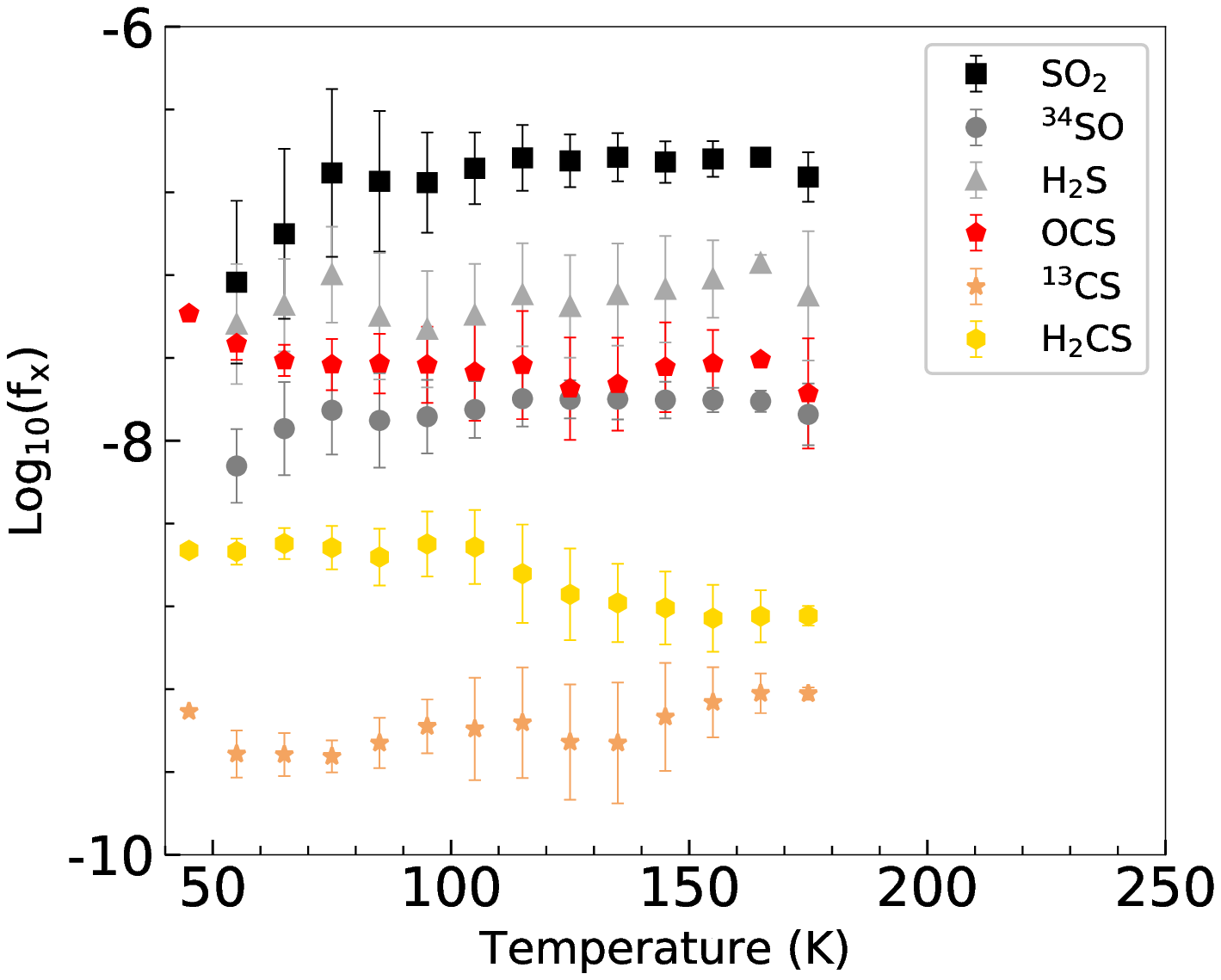}
\caption{The correlation between the gas temperature (here, we take the rotation temperature of $\rm SO_2$) and molecular relative abundance ratios with respect to $\rm H_2$ (in different colors). Each point shows the statistical mean in a temperature bin of 10 K.\label{fig:T_ab}}
\end{figure}

Inspecting possible correlations between both parameters pixel by pixel (Figure \ref{fig:T_ab}), we note that the abundances of OCS and $^{13}$CS decrease as the gas temperature increases from 40\,K to 180\,K, and that of H$_2$CS seems not sensitive to this gas temperature range. As for the carbon-free S-bearing species, their abundances exhibit a drastic increase as the gas temperature changes from 40\,K to 90\,K, increasing by an order of magnitude. Similar observational results confirmed that the abundance of SO$_2$ was enhanced by more than two orders of magnitude from cold outer envelopes (T $<$ 100\,K) to hot inner envelopes (T $>$ 100\,K) \citep{Van2003}. However, molecular abundances of these species exhibit no variation when the gas temperature increase from 90\,K to 180\,K, which may be the result of depletion mechanisms \citep{Wakelam2011}.

\subsection{Possible chemical relation}\label{sec:relative abundance}
It has long been proposed that relative abundance ratios of different S-bearing species have the possibility to trace the chemical history of star-forming regions \citep [e.g.,][]{Charnley1997, Hatchell1998, Wakelam2004, Wakelam2011}. In Orion KL, the abundances of two groups of S-bearing species seem to be dependent on the gas temperature (Section~\ref{sec:abundance}). Therefore, we investigate the possible correlation between the gas temperature and the relative abundance ratios of $\mathrm{^{34}SO/^{34}SO_2}$, $\mathrm{OCS/SO_2}$, and $\mathrm{H_2S/SO_2}$ (Figure \ref{fig:relative abundance}).

The $\mathrm{^{34}SO/^{34}SO_2}$ ratio decreases from 1.5 to 0.7 as the temperature increases from 60$\sim$180 K, which is consistent with the results of \citep{Esplugues2014}. It is known that SO can easily convert to SO$_2$ as the temperature increases \citep{Charnley1997}. In particular, shocks can enhance the abundances of SO and SO$_2$ by two orders of magnitudes \citep[e.g.,][]{Pineau1993,Bachiller1997,Wright2017}. As shown in Section \ref{sec:gas distribution}, both SO and SO$_2$ may enhance by shock events in Orion KL. 

The $\mathrm{OCS/SO_2}$ ratio decreases from 1.0 to 0.3 when the temperature increases from 40\,K to 100\,K and stays constant at $\sim$0.2 from 100\,K to 180\,K, consistent with the model prediction by \citealt{Wakelam2011} (shown as their Figure~7). SO$_2$ may be produced from species such as SO or evaporated to the gas phase at $>100\,K$, whereas OCS may be destructed in such temperature regime.

The $\mathrm{H_2S/SO_2}$ ratio is $\rm\sim$0.3, exhibiting no obvious variations in temperature range of 50-180\,K. This result is consistent with that of \citealt{Esplugues2014}, which indicates a chemical age of $\sim$5$\times$10$^4$ years for the hot core region.

\begin{figure}
  \includegraphics[width=1.0\linewidth,trim=0 0 0 0,clip]{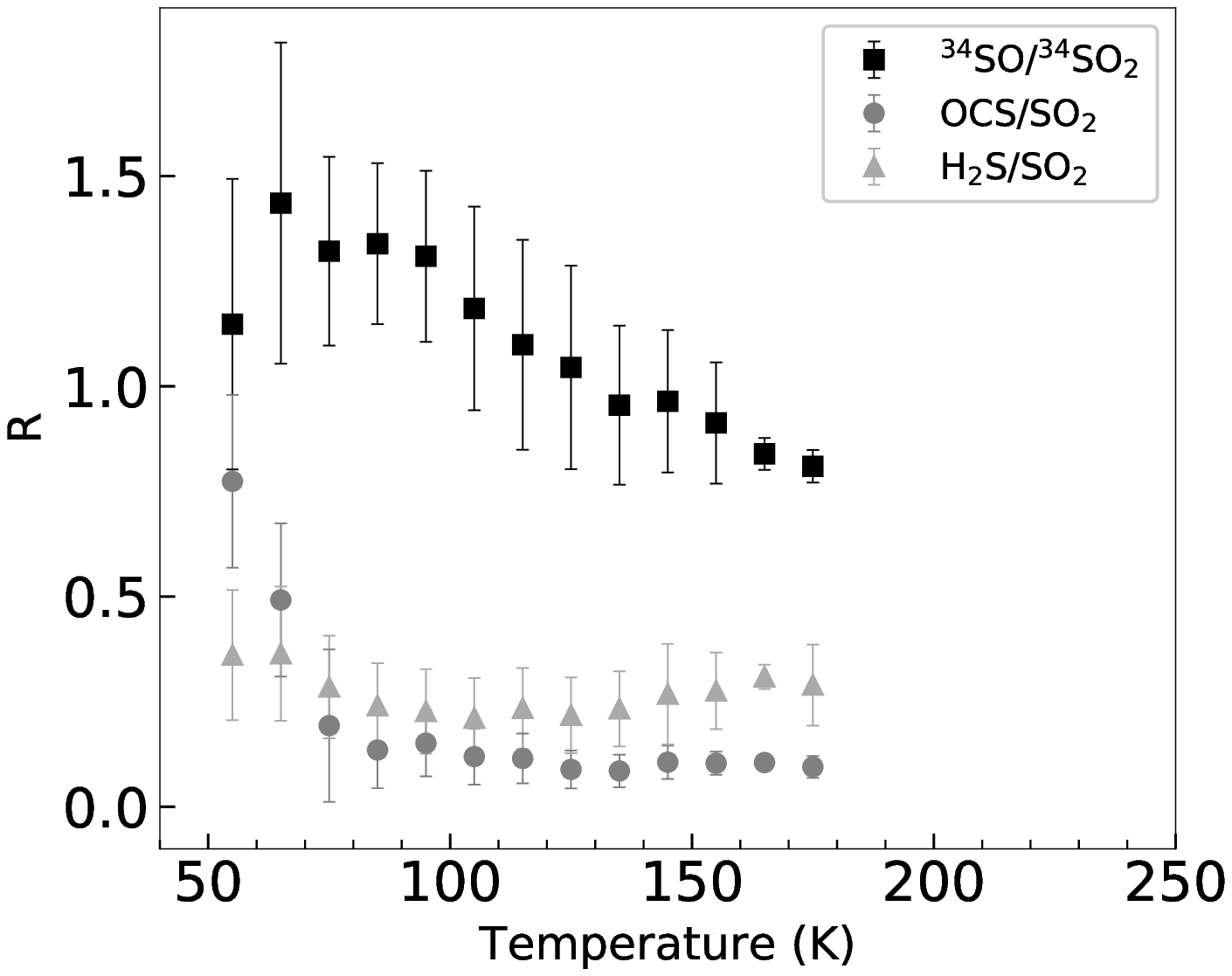}
        \caption{The relative abundance ratios of $\mathrm{^{34}SO/^{34}SO_2}$, $\mathrm{OCS/SO_2}$, and $\mathrm{H_2S/SO_2}$ at different temperatures. Each point shows the statistical mean in a temperature bin of 10 K.}
         \label{fig:relative abundance}
\end{figure}

\section{Conclusion}\label{sec:conclusion}
We constructed a combined ALMA and IRAM-30\,m data set around the 1.3\,mm band with a linear resolution of $\sim$800\,au and a velocity resolution of 1\,\kms, in which 79 molecular lines from six S-bearing species (SO$_2$, SO, H$_2$S, OCS, $\mathrm{^{13}CS}$, H$_2$CS) were identified.
A clear dichotomy was found between carbon-sulfur compounds (OCS, $^{13}$CS, H$_2$CS) and carbon-free S-bearing species (SO$_2$, SO, H$_2$S), in terms of their spatial distributions and kinematic features. The main conclusions are as follows:

1. Using the XCLASS package, we fit the synthetic spectrum of each species pixel by pixel and derive the rotational temperature map of SO$_2$ under LTE. A gradient from the warmest hot core ($\sim$176\,K) to the surrounding substructures (mm2, mm3, mm4, and mm5, $\sim$106\,K) and the southern region mm7  ($\sim$68\,K) was found.

2. The carbon-sulfur compounds (i.e., OCS, $^{13}$CS, H$_2$CS) exhibit spatial distributions concentrated towards the continuum peaks and extended to the south ridge. The carbon-free S-bearing species extended to the northeast of mm4. Specifically, there is a ring-like structure appears in the channel maps of the SO$_2$, $^{34}$SO, and H$_2$S lines at velocity from 10 to 15\,\kms, which may be influenced by shocks from the OMC1 explosion.

3. The FWHM linewidths of carbon-sulfur compounds are in the range of 2 $\sim$ 11 \kms, increasing as temperature increase. The carbon-free S-bearing species exhibit broader FWHM linewidths (12 $\sim$ 26 \kms) towards mm2, mm3, mm4, and mm5, which is significant broader than hot core (7--14 \kms) and mm7 (3--8 \kms).

4. The molecular abundances of OCS and H$_2$CS with respect to $\rm H_2$ decrease from the cold ($\sim$68 K) to the hot ($\sim$ 176\,K) regions. In contrast, the molecular abundances of carbon-free S-bearing species with respect to $\rm H_2$ increase by an order of magnitude when the temperature increases from 50\,K to 100\,K.

5. The relative abundance ratios of $\mathrm{^{34}SO/^{34}SO_2}$ and $\mathrm{OCS/SO_2}$ enhanced in the warmer regions ($>$100\,K) with respect to the colder regions ($\sim$50\,K). Such enhancements consistent with the transformation of SO$_2$ at warmer regions and the influence of shocks.

\acknowledgments
This work has been supported by the National Key R\&D Program of China No.\ 2017YFA0402600, the CAS Strategic Priority Research Program No.\ XDB23000000, the CAS International Partnership Program No.\ 114A11KYSB20160008, and the National Natural Science Foundation of China No.\ 11725313.

SF acknowledges the support of the EACOA fellowship from the East Asia Core Observatories Association, which consists of the National Astronomical Observatory of China, the National Astronomical Observatory of Japan, the Academia Sinica Institute of Astronomy and Astrophysics, and the Korea Astronomy and Space Science Institute.

SLQ is supported by the Joint Research Fund in Astronomy (U1631237) under a cooperative agreement between the National Natural Science Foundation of China (NSFC) and the Chinese Academy of Sciences (CAS) and by the Top Talents Program of Yunnan Province (2015HA030).

Zhiyuan Ren is supported by National Natural Science Foundation of China (U1731218).

This paper makes use of the following ALMA data: ADS/ JAO.ALMA\#2011.0.00009.SV. ALMA is a partnership of ESO (representing its member states), NSF (USA), and NINS (Japan), together with NRC (Canada), NSC and ASIAA (Taiwan), and KASI (Republic of Korea) in cooperation with the Republic of Chile. The Joint ALMA Observatory is operated by ESO, AUI/NRAO, and NAOJ.

\vspace{5mm}

\software{MIRIAD \citep{miriad},
          GILDAS/CLASS \citep{Pety2005,Gildas2013},
          XCLASS \citep{xclass},
          MAGIX \citep{magix}
          }
\bibliography{reference}
\clearpage

\appendix
\setcounter{table}{0}
\renewcommand{\thetable}{A\arabic{table}}
\section{Appendix 1: Transitions and fitting results} \label{sec:appendix}
\startlongtable
\begin{deluxetable*}{clcrrcccc}
\tablecaption{All the combined spectral lines identified in this paper\label{tab:spectral_lines} }
\tablewidth{700pt}
\tabletypesize{\scriptsize}
\tablehead{
\colhead{Species} & \colhead{Transition} & \colhead{Rest Frequency} & \colhead{E$\mathrm{_u}$} & \colhead{S$\mathrm{_{ij}\mu^{2}}$} & \colhead{Log$\mathrm{_{10}(A_{ij})}$} & \colhead{Blended} & \colhead{RMS} & \colhead{$\mathrm{I_{peak}}$(hot core)} \\
 &  & \colhead{(MHz)} & \colhead{(K)} &  &  & & \colhead{\jbeam}& \colhead{\jbeam}
}
\startdata
  & $16_{3,13}-16_{2,14}$ & 214689.4 & 147.83 & 28.37 & -4.00 & &0.11&11.0\\
   & $17_{6,12}-18_{5,13}$ & 214728.3 & 228.96 & 5.75 & -4.72 & &0.05&7.14 \\
   & $26_{2,24}-27_{1,27}$ & 215094.5 & 340.55 & 0.41 & -6.05 &  $\mathrm{CH_3CH_2CN}$& 0.06 & \\
   & $22_{2,20}-22_{1,21}$ & 216643.3 & 248.44 & 35.25 & -4.03 & &0.08&10.5 \\
   & $22_{7,15}-23_{6,18}$ & 219276.0 & 352.75 & 7.81 & -4.67 & &0.05&5.56 \\
   & $36_{3,33}-37_{2,36}$ & 222869.1 & 648.60 & 0.96 & -5.77 & \tablenotemark{*}&0.004 &0.27\\
   & $27_{8,20}-28_{7,21}$ & 223434.5 & 504.42 & 9.87 & -4.63 & \tablenotemark{*}& 0.01 &2.61\\
   & $46_{4,42}-47_{3,45}$ & 224473.4 & 1054.52 & 1.59 & -5.65 & blended with $\mathrm{CH_3CH_2CN}$\tablenotemark{*}& 0.04& \\
   & $41_{5,37}-40_{6,34}$ & 226508.3 & 856.96 & 15.25 & -4.60 & \tablenotemark{*}& 0.007& 0.64\\
   & $32_{9,23}-33_{8,26}$ & 227335.8 & 683.92 & 11.93 & -4.60 & \tablenotemark{*}& 0.006&1.21\\
   & $11_{5,7}-12_{4,8}$ & 229347.6 & 122.00 & 3.13 & -4.72 & &0.05 &8.80 \\
 $\mathrm{SO_2}$, $\upsilon$$\mathrm{=0}$  & $45_{6,40}-44_{7,37}$ & 229749.7 & 1044.76 & 18.38 & -4.55 & $\mathrm{CH_3OH}$& 0.08 & \\
   & $37_{10,28}-38_{9,29}$ & 230965.2 & 891.24 & 13.98 & -4.57 & & 0.04 & 0.83 \\
   & $28_{3,25}-28_{2,26}$ & 234187.1 & 403.03 & 55.10 & -3.84 & & 0.06 & 9.50 \\
   & $42_{11,31}-43_{10,34}$ & 234353.0 & 1126.34 & 16.04 & -4.55 & $\mathrm{Na^{13}CN/NaN^{13}C}$ & 0.03& \\
   & $16_{6,10}-17_{5,13}$ & 234421.6 & 213.32 & 5.17 & -4.63 & $\mathrm{CH_3CH_2CN}$ & 0.07&\\
   & $4_{2,2}-3_{1,3}$ & 235151.7 & 19.03 & 4.57 & -4.11 & & 0.10 & 9.68 \\
   & $16_{1,15}-15_{2,14}$ & 236216.7 & 130.66 & 16.14 & -4.12 & & 0.19 & 11.7 \\
  \cline{2-9}
   & $11_{5,7}-12_{4,8}$ & 213807.3 & 119.95 & 3.12 & -4.81 & & 0.03 & 0.83 \\
   & $16_{6,10}-17_{5,13}$ & 215468.1 & 210.34 & 5.17 & -4.74 & blended & 0.05 &\\
   & $14_{3,11}-14_{2,12}$ & 215999.7 & 118.26 & 23.32 & -4.03 & &0.05 &4.42  \\
   & $21_{7,15}-22_{6,16}$ & 216593.5 & 328.45 & 7.23 & -4.70 & & 0.04 & 1.70  \\
   & $11_{1,11}-10_{0,10}$ & 219355.0 & 60.16 & 20.75 & -3.96 & & 0.05 & 5.94 \\
$\mathrm{^{34}SO_2}$   & $22_{2,20}-22_{1,21}$ & 221114.9 & 248.19 & 33.77 & -4.02 & blended & 0.03 & \\
   & $4_{2,2}-3_{1,3}$ & 229857.6 & 18.70 & 4.54 & -4.15 &   $\mathrm{CH_3OH,CH_3CHO,OS^{18}O}$ & 0.04 & \\
   & $5_{4,2}-6_{3,3}$ & 230933.4 & 51.76 & 0.68 & -5.06 & & 0.04 & 0.57 \\
   & $15_{6,10}-16_{5,11}$ & 235004.0 & 195.63 & 4.59 & -4.56 & $\mathrm{(CH_3)_2CO}$ & 0.02 & \\
   & $5_{2,4}-4_{1,3}$ & 235927.5 & 23.26 & 5.66 & -4.10 & blended & 0.06 &\\
   & $10_{3,7}-10_{2,8}$ & 235951.9 & 71.97 & 14.60 & -3.97 & blended & 0.06 &\\
   & $20_{2,18}-19_{3,17}$ & 236225.1 & 207.54 & 12.70 & -4.32 & $\mathrm{SO_2}$ & 0.19 &\\
   & $20_{7,13}-21_{6,16}$ & 236295.7 & 309.13 & 6.64 & -4.60 & & 0.03 & 0.84 \\
   & $19_{3,17}-20_{0,20}$ & 236428.8 & 196.18 & 0.35 & -5.86 & $\mathrm{CH_2CH^{13}CN}$ & 0.03 & \\
  \cline{2-9}
   & $6_{4,2}-7_{3,5}$,$\mathrm{F=11/2-11/2}$ & 217628.2 & 58.71 & 0.03 & -6.46 & & & \\
   & $6_{4,2}-7_{3,5}$,$\mathrm{F=9/2-11/2}$ & 217628.2 & 58.71 & 0.88 & -4.98 & & &  \\
   $\mathrm{^{33}SO_2}$ & $6_{4,2}-7_{3,5}$,$\mathrm{F=15/2-17/2}$ & 217628.4 & 58.71 & 1.37 & -4.99 & & 0.04 & 0.23 \\
   & $6_{4,2}-7_{3,5}$,$\mathrm{F=11/2-13/2}$ & 217628.7 & 58.71 & 1.02 & -4.99 & & &  \\
   & $6_{4,2}-7_{3,5}$,$\mathrm{F=13/2-13/2}$ & 217628.8 & 58.71 & 0.05 & -6.40 & & & \\
   & $6_{4,2}-7_{3,5}$,$\mathrm{F=13/2-15/2}$ & 217628.9 & 58.71 & 1.18 & -4.99 & & & \\
   & $6_{4,2}-7_{3,5}$,$\mathrm{F=15/2-15/2}$ & 217629.1 & 58.71 & 0.03 & -6.58 & & & \\
   & $22_{2,20}-22_{1,21}$,$\mathrm{F=45/2-45/2}$ & 218875.4 & 251.78 & 34.93 & -4.03 & & &\\
   & $22_{2,20}-22_{1,21}$,$\mathrm{F=45/2-43/2}$ & 218875.6 & 251.78 & 0.14 & -6.44 & & & \\
   & $22_{2,20}-22_{1,21}$,$\mathrm{F=43/2-45/2}$ & 218875.7 & 251.78 & 0.14 & -6.42 & & & \\
   & $22_{2,20}-22_{1,21}$,$\mathrm{F=43/2-43/2}$ & 218875.9 & 251.78 & 33.41 & -4.03 &  & &  \\
   & $22_{2,20}-22_{1,21}$,$\mathrm{F=45/2-47/2}$ & 218876.6 & 251.78 & 0.10 & -6.57 & & &    \\
   & $22_{2,20}-22_{1,21}$,$\mathrm{F=43/2-41/2}$ & 218877.3 & 251.78 & 0.10 & -6.55 &  & 0.05& 1.00 \\
   & $22_{2,20}-22_{1,21}$,$\mathrm{F=47/2-45/2}$ & 218880.7 & 251.78 & 0.10 & -6.59 &  & &  \\
   & $22_{2,20}-22_{1,21}$,$\mathrm{F=41/2-43/2}$ & 218880.9 & 251.78 & 0.10 & -6.53 & & &   \\
   & $22_{2,20}-22_{1,21}$,$\mathrm{F=47/2-47/2}$ & 218881.9 & 251.78 & 36.60 & -4.03 & & &   \\
   & $22_{2,20}-22_{1,21}$,$\mathrm{F=41/2-41/2}$ & 218882.3 & 251.78 & 32.01 & -4.03 & & &   \\
   & $11_{1,11}-10_{0,10}$,$\mathrm{F=19/2-19/2}$ & 220613.4 & 61.10 & 0.25 & -5.80 & & &  \\
   & $11_{1,11}-10_{0,10}$,$\mathrm{F=19/2-17/2}$ & 220617.4 & 61.10 & 17.51 & -3.96 & & &   \\
   & $11_{1,11}-10_{0,10}$,$\mathrm{F=25/2-23/2}$ & 220617.8 & 61.10 & 23.09 & -3.95 & $\mathrm{CH_3^{13}CN}$ &0.05 & \\
   & $11_{1,11}-10_{0,10}$,$\mathrm{F=21/2-21/2}$ & 220619.7 & 61.10 & 0.34 & -5.72 & & &   \\
   & $11_{1,11}-10_{0,10}$,$\mathrm{F=21/2-19/2}$ & 220620.4 & 61.10 & 19.20 & -3.96 & & &   \\
   & $11_{1,11}-10_{0,10}$,$\mathrm{F=23/2-21/2}$ & 220620.7 & 61.10 & 21.06 & -3.96 & & &   \\
   & $11_{1,11}-10_{0,10}$,$\mathrm{F=23/2-23/2}$ & 220624.7 & 61.10 & 0.25 & -5.88 & & &   \\
   & $14_{3,11}-14_{2,12}$,$\mathrm{F=27/2-25/2}$ & 220983.1 & 61.10 & 0.16 & -6.14 & & &  \\
   & $14_{3,11}-14_{2,12}$,$\mathrm{F=29/2-31/2}$ & 220983.3 & 61.10 & 0.16 & -6.17 & & &   \\
   & $14_{3,11}-14_{2,12}$,$\mathrm{F=25/2-25/2}$ & 220985.4 & 61.10 & 20.40 & -4.01 & & &   \\
   & $14_{3,11}-14_{2,12}$,$\mathrm{F=31/2-31/2}$ & 220985.8 & 61.10 & 25.15 & -4.01 & & &   \\
   $\mathrm{^{33}SO_2}$  & $14_{3,11}-14_{2,12}$,$\mathrm{F=29/2-27/2}$ & 220988.5 & 120.26 & 0.22 & -6.04 &blended  &0.03 &  \\
   & $14_{3,11}-14_{2,12}$,$\mathrm{F=27/2-27/2}$ & 220988.7 & 120.26 & 21.77 & -4.01 & & &   \\
   & $14_{3,11}-14_{2,12}$,$\mathrm{F=29/2-29/2}$ & 220989.0 & 120.26 & 23.35 & -4.01 & & &   \\
   & $14_{3,11}-14_{2,12}$,$\mathrm{F=27/2-29/2}$ & 220989.2 & 120.26 & 0.22 & -6.01 & & &   \\
   & $14_{3,11}-14_{2,12}$,$\mathrm{F=25/2-27/2}$ & 220991.0 & 120.26 & 0.16 & -6.10 & & &   \\
   & $14_{3,11}-14_{2,12}$,$\mathrm{F=31/2-29/2}$ & 220991.5 & 120.26 & 0.16 & -6.19 & & &   \\
  &$20_{2,18}-19_{3,17}$,$\mathrm{F=41/2-41/2}$&230435.3&210.54&0.05&-6.80& & & \\
  &$20_{2,18}-19_{3,17}$,$\mathrm{F=41/2-39/2}$&230436.4&210.54&12.71&-4.37& & &  \\
  &$20_{2,18}-19_{3,17}$,$\mathrm{F=39/2-39/2}$&230436.7&210.54&0.06&-6.65& & &  \\
  &$20_{2,18}-19_{3,17}$,$\mathrm{F=39/2-37/2}$&230436.7&210.54&12.08&-4.37& &0.03 &0.55\\
  &$20_{2,18}-19_{3,17}$,$\mathrm{F=43/2-41/2}$&230441.0&210.54&13.36&-4.36& & &  \\
  &$20_{2,18}-19_{3,17}$,$\mathrm{F=37/2-35/2}$&230441.4&210.54&11.49&-4.37& & &  \\
  &$20_{2,18}-19_{3,17}$,$\mathrm{F=37/2-37/2}$&230442.2&210.54&0.05&-6.75& & &  \\
  &$12_{3,9}-12_{2,10}$,$\mathrm{F=23/2-21/2}$&231894.9&94.88&0.18&-5.98& & & \\
  &$12_{3,9}-12_{2,10}$,$\mathrm{F=25/2-27/2}$&231895.1&94.88&0.18&-6.01& & &  \\
  &$12_{3,9}-12_{2,10}$,$\mathrm{F=21/2-21/2}$&231896.6&94.88&15.94&-3.98& & &  \\
  &$12_{3,9}-12_{2,10}$,$\mathrm{F=27/2-27/2}$&231897.0&94.88&20.33&-3.98& & &  \\
  &$12_{3,9}-12_{2,10}$,$\mathrm{F=25/2-23/2}$&231899.7&94.88&0.23&-5.89&$\mathrm{CH_3C^{15}N}$, $\mathrm{C_2H_5^{13}CN}$ &0.05 &  \\
  &$12_{3,9}-12_{2,10}$,$\mathrm{F=23/2-23/2}$&231899.8&94.88&17.17&-3.98& & &  \\
  &$12_{3,9}-12_{2,10}$,$\mathrm{F=25/2-25/2}$&231900.2&94.88&18.63&-3.98& & &  \\
  &$12_{3,9}-12_{2,10}$,$\mathrm{F=23/2-25/2}$&231900.4&94.88&0.23&-5.85& & &  \\
  &$12_{3,9}-12_{2,10}$,$\mathrm{F=21/2-23/2}$&231901.5&94.88&0.18&-5.94& & &  \\
  &$12_{3,9}-12_{2,10}$,$\mathrm{F=27/2-25/2}$&231902.1&94.88&0.18&-6.04& & &  \\
  &$4_{2,2}-3_{1,3}$,$\mathrm{F=9/2-7/2}$&232415.3&19.12&4.58&-4.17& & &  \\
  &$4_{2,2}-3_{1,3}$,$\mathrm{F=7/2-7/2}$&232415.6&19.12&0.54&-5.00& & &  \\
  &$4_{2,2}-3_{1,3}$,$\mathrm{F=5/2-7/2}$&232416.5&19.12&0.02&-6.43& & &  \\
  &$4_{2,2}-3_{1,3}$,$\mathrm{F=7/2-5/2}$&232418.4&19.12&3.44&-4.20&$\mathrm{CH_3OH}$ &0.10 &  \\
  &$4_{2,2}-3_{1,3}$,$\mathrm{F=5/2-5/2}$&232419.3&19.12&0.41&-5.00& & &  \\
  $\mathrm{^{33}SO_2}$ &$4_{2,2}-3_{1,3}$,$\mathrm{F=9/2-9/2}$&232421.1&19.12&0.42&-5.22& & &  \\
  &$4_{2,2}-3_{1,3}$,$\mathrm{F=11/2-9/2}$&232422.2&19.12&6.00&-4.14& & &  \\
  &$4_{2,2}-3_{1,3}$,$\mathrm{F=5/2-3/2}$&232425.2&19.12&2.57&-4.20& & &  \\
  &$28_{3,25}-28_{2,26}$,$\mathrm{F=57/2-59/2}$&235722.4&408.45&0.10&-6.58& & & \\
  &$28_{3,25}-28_{2,26}$,$\mathrm{F=55/2-53/2}$&235722.9&408.45&0.10&-6.56& & &  \\
  &$28_{3,25}-28_{2,26}$,$\mathrm{F=57/2-57/2}$&235724.9&408.45&55.16&-3.84& & &  \\
  &$28_{3,25}-28_{2,26}$,$\mathrm{F=57/2-55/2}$&235724.9&408.45&0.13&-6.45& & &  \\
  &$28_{3,25}-28_{2,26}$,$\mathrm{F=55/2-57/2}$&235725.1&408.45&0.13&-6.44& & &   \\
  &$28_{3,25}-28_{2,26}$,$\mathrm{F=55/2-55/2}$&235725.1&408.45&53.25&-3.84& &0.04 &0.75  \\
  &$28_{3,25}-28_{2,26}$,$\mathrm{F=59/2-59/2}$&235728.7&408.45&57.20&-3.84& & &  \\
  &$28_{3,25}-28_{2,26}$,$\mathrm{F=53/2-53/2}$&235728.9&408.45&51.46&-3.84& & &  \\
  &$28_{3,25}-28_{2,26}$,$\mathrm{F=53/2-55/2}$&235731.1&408.45&0.10&-6.55& & &  \\
  &$28_{3,25}-28_{2,26}$,$\mathrm{F=59/2-57/2}$&235731.2&408.45&0.10&-6.59& & &  \\
  \cline{2-9}
  &$27_{3,24}-27_{2,25}$&215587.1&377.80&336.41&-2.88&blended &0.08 & \\
  &$13_{0,13}-12_{1,12}$&217018.4&81.94&146.71&-3.55& $\mathrm{(CH_3)_2CO}$ &0.05 & \\
  &$15_{3,12}-15_{2,13}$&219240.2&133.72&148.94&-3.67&$\mathrm{(CH_3)_2CO}$ &0.03 & \\
  &$13_{3,10}-13_{2,11}$&230231.4&106.72&119.90&-3.56&$\mathrm{C_3H_8}$ &0.04 & \\
  $\mathrm{OS^{17}O}$  &$11_{5,7}-12_{4,8}$&230569.4&123.49&18.37&-4.16&CO &0.17 & \\
  &$12_{1,12}-11_{0,11}$&230684.3&71.17&134.20&-3.42& & 0.04&0.30  \\
  &$14_{2,13}-14_{1,14}$&230810.9&106.25&81.07&-3.80&$\mathrm{CH_2CHCN}$ &0.04 & \\
  &$18_{1,17}-18_{0,18}$&232123.6&163.68&96.08&-3.20&$\mathrm{^{13}CH_3CN}$ &0.03 & \\
  &$14_{0,14}-13_{1,13}$&236758.0&94.30&163.17&-3.47&blended &0.04 & \\
  \cline{2-9}
  &$13_{2,12}-13_{1,13}$&215756.4&92.58&13.01&-4.25&blended &0.04 & \\
  &$23_{2,21}-23_{1,22}$&216415.5&268.02&35.95&-4.04& &0.04 &0.419  \\
  $\mathrm{OS^{18}O}$   &$16_{1,15}-15_{2,14}$&217102.7&129.67&15.17&-4.26&$\mathrm{SiO}$ &0.12 & \\
  &$4_{2,3}-3_{1,2}$&218230.3&19.19&4.79&-4.19&blended &0.09 & \\
  &$15_{3,12}-15_{2,13}$&218316.9&132.52&24.36&-4.02&$\mathrm{CH_2CHCN}$ &0.10 & \\
  &$14_{0,14}-13_{1,13}$&229854.9&93.29&26.72&-3.89&$\mathrm{^{34}SO_2}$ &0.04 & \\
  &$5_{2,4}-4_{1,3}$&233497.6&23.73&5.41&-4.14&$\mathrm{CH_3CH_2CN}$ &0.05 & \\
  $\mathrm{OS^{18}O}$   &$12_{3,9}-12_{2,10}$&233588.4&93.92&17.57&-3.98&$\mathrm{CH_2CDCN}$ &0.04 & \\
  &$24_{2,22}-24_{1,23}$&233950.0&290.45&35.43&-3.97&blended &0.03 & \\
  &$15_{2,14}-15_{1,15}$&236805.0&118.73&13.90&-4.15&$\mathrm{^{13}CH_2CHCN}$ &0.06 & \\
  \cline{2-9}
  &$7_8-7_7$&214357.0&81.24&0.44&-5.47&$\mathrm{^{13}CH_3CN}$ &0.08 & \\
  SO  &$5_5-4_4$&215220.7&44.10&11.31&-3.92& &0.19 &15.2  \\
  &$6_5-5_4$&219949.4&34.98&14.01&-3.87& &0.17 &16.6  \\
  &$1_2-2_1$&236452.3&15.81&0.03&-5.85& &0.03 &1.60  \\
  \cline{2-9}
  $^{34}$SO&$6_5-5_4$&215839.9&34.38&14.02&-3.90& &0.09 &9.58  \\
  \cline{2-9}
  &$6_5-5_4$,$\mathrm{F=9/2-7/2}$&217827.2&34.67&10.20&-3.91& & & \\
  $\mathrm{^{33}SO}$  &$6_5-5_4$,$\mathrm{F=11/2-9/2}$&217829.8&34.67&12.17&-3.91&blended &0.06 &  \\
  &$6_5-5_4$,$\mathrm{F=13/2-11/2}$&217831.8&34.68&14.52&-3.90& & &  \\
  &$6_5-5_4$,$\mathrm{F=15/2-13/2}$&217832.6&34.68&17.26&-3.89& & &  \\
  \cline{2-9}
  $\mathrm{S^{18}O}$&$5_6-4_5$&232265.8&47.76&11.38&-3.82&$\mathrm{^{39}SiC_2}$ &0.04 & \\
  \cline{2-9}
  \multirow{2}{*}{OCS}
  &$18-17$&218903.4&99.81&9.21&-4.52& &0.09 &9.78  \\
  &$19-18$&231061.0&110.90&9.72&-4.44& &0.10 &9.40  \\
  \cline{2-9}
  $\mathrm{OC^{33}S}$&$18-17$&216147.4&98.55&9.21&-4.04&$\mathrm{NH_2CH_2CN}$ &0.04 & \\
  \cline{2-9}
  $\mathrm{O^{13}CS}$
  &$18-17$&218199.0&99.49&9.21&-4.52& &0.05 &1.53  \\
  &$19-18$&230317.5&110.54&9.72&-4.45& &0.04 &1.34  \\
  \cline{2-9}
  $\mathrm{^{18}OCS}$&$19-18$&216753.5&104.04&9.70&-4.53&$\mathrm{CH_3CH_2CN}$ &0.04 & \\
  \cline{2-9}
  $\mathrm{^{13}CS}$&$5-4$&231220.7&33.29&38.33&-3.60& &0.05 &5.86  \\
  \cline{2-9}
  $\mathrm{H_2S}$&$2_{2,0}-2_{1,1}$&216710.4&83.98&2.06&-4.31& &0.09 &8.89  \\
  \cline{2-9}
  $\mathrm{H_2^{34}S}$&$2_{2,0}-2_{1,1}$&214376.9&83.80&2.07&-4.32&$\mathrm{^{13}CH_3CN}$ &0.09 & \\
  \cline{2-9}
  &$2_{2,0}-2_{1,1}$,$\mathrm{J=1/2-3/2}$&215494.4&83.89&0.41&-4.62& & &  \\
  &$2_{2,0}-2_{1,1}$,$\mathrm{J=1/2-1/2}$&215496.7&83.89&0.41&-4.62& & &  \\
  &$2_{2,0}-2_{1,1}$,$\mathrm{J=7/2-5/2}$&215500.8&83.89&0.47&-5.16& & &  \\
  &$2_{2,0}-2_{1,1}$,$\mathrm{J=7/2-7/2}$&215502.8&83.89&2.83&-4.39& & &  \\
  $\mathrm{H_2^{33}S}$  &$2_{2,0}-2_{1,1}$,$\mathrm{J=3/2-5/2}$&215503.8&83.89&0.58&-4.77&blended &0.06 &  \\
  &$2_{2,0}-2_{1,1}$,$\mathrm{J=3/2-3/2}$&215505.4&83.89&0.66&-4.72& & &  \\
  &$2_{2,0}-2_{1,1}$,$\mathrm{J=3/2-1/2}$&215507.6&83.89&0.41&-4.92& & &  \\
  &$2_{2,0}-2_{1,1}$,$\mathrm{J=5/2-5/2}$&215511.6&83.89&1.42&-4.56& & &  \\
  &$2_{2,0}-2_{1,1}$,$\mathrm{J=5/2-3/2}$&215513.4&83.89&0.58&-4.95& & &  \\
  &$2_{2,0}-2_{1,1}$,$\mathrm{J=5/2-7/2}$&215513.4&83.89&0.47&-5.04& & &  \\
  \cline{2-9}
  $\mathrm{H_2CS}$&$7_{1,7}-6_{1,6}$&236727.0&58.62&55.95&-3.72& &0.05 &2.99  \\
  \cline{2-9}
  $\mathrm{H_2C^{34}S}$&$7_{1,7}-6_{1,6}$&232754.7&57.87&55.89&-3.74& &0.04 &0.61  \\
  &$7_{2,5}-6_{2,4}$&236441.8&98.11&17.47&-3.75& &0.02 &0.27  \\
  \cline{2-9}
  &$7_{1,7}-6_{1,6}$,$\mathrm{J=11/2-11/2}$&234670.6&58.23&1.70&-5.15& & & \\
  &$7_{1,7}-6_{1,6}$,$\mathrm{J=13/2-13/2}$&234677.0&58.23&2.26&-5.09& & &  \\
  &$7_{1,7}-6_{1,6}$,$\mathrm{J=15/2-13/2}$&234678.8&58.23&57.92&-3.74& & &  \\
  $\mathrm{H_2C^{33}S}$  &$7_{1,7}-6_{1,6}$,$\mathrm{J=17/2-15/2}$&234678.8&58.23&67.07&-3.73&$\mathrm{CH_3OH}$ &0.07 &  \\
  &$7_{1,7}-6_{1,6}$,$\mathrm{J=13/2-11/2}$&234679.0&58.23&49.89&-3.75& & &  \\
  &$7_{1,7}-6_{1,6}$,$\mathrm{J=11/2-9/2}$&234679.1&58.23&42.99&-3.75& & &  \\
  &$7_{1,7}-6_{1,6}$,$\mathrm{J=15/2-15/2}$&234687.0&58.23&1.70&-5.27& & &  \\
  \cline{2-9}
\enddata
\tablenotetext{*}{Lines from ALMA alone.}
\end{deluxetable*}

\startlongtable
\begin{deluxetable}{ccrrr}
\tablecaption{The optimum solutions of different parameters.\label{tab:parameters}}
\tablewidth{700pt}
\tabletypesize{\scriptsize}
\tablehead{
\colhead{Species} & \colhead{T$\mathrm{_{rot}}$} & \colhead{Column density} & \colhead{$\Delta$ V} & \colhead{V$\mathrm{_{lsr}}$} \\
 & \colhead{(K)} & \colhead{($\mathrm{\times 10^{16} cm^{-2}}$)} & \colhead{(\kms)} & \colhead{(\kms)}
}
\startdata
  \multicolumn{5}{c}{hot core}\\
  \hline
  $\mathrm{SO_2}$& \multirow{14}{*}{$176 \pm 9$} &$58 \pm 12$&$6.7 \pm 0.1$&\multirow{5}{*}{$7.8 \pm 0.1$}\\
  $\mathrm{^{34}SO_2}$& &$6 \pm 1$&$9.8 \pm 0.3$& \\
  $\mathrm{^{33}SO_2}$& &0.2&3.00& \\
  $\mathrm{OS^{17}O}$& &$0.08 \pm 0.03$&$5 \pm 3$& \\
  $\mathrm{OS^{18}O}$& &$0.20$&$2.00$& \\
  $\mathrm{^{34}SO}$& &$4 \pm 2$&$12.6 \pm 0.5$&\multirow{3}{*}{$7.6 \pm 0.2$}\\
  $\mathrm{^{33}SO}$& &$1.3 \pm 0.6$&$12.4 \pm 0.5$& \\
  OCS& &$10 \pm 6$&$11.0 \pm 0.5$&\multirow{3}{*}{$7.0 \pm 0.2$}\\
  $\mathrm{O^{13}CS}$& &$0.5 \pm 0.2$&$7.2 \pm 0.5$& \\
  $\mathrm{^{13}CS}$& &$0.3 \pm 0.1$&$7.5 \pm 0.4$&$7.0 \pm 0.2$\\
  $\mathrm{H_2S}$& &$17 \pm 7$&$13.9 \pm 0.6$&$7.1 \pm 0.2$\\
  $\mathrm{H_2CS}$& &$0.8 \pm 0.3$&$7.1 \pm 0.3$&\multirow{2}{*}{$8.2 \pm 0.1$}\\
  $\mathrm{H_2C^{34}S}$& &$0.042 \pm 0.003$&$4 \pm 4$& \\
  $\mathrm{C^{18}O}$& &$40 \pm 19$&$9.4 \pm 0.3$&$7.7 \pm 0.2$\\
  \hline
  \multicolumn{5}{c}{mm2}\\
  \hline
  $\mathrm{SO_2}$& \multirow{14}{*}{$105 \pm 7$} &$31 \pm 14 $&$17.9 \pm 0.5$&\multirow{5}{*}{$6.5 \pm 0.3$}\\
  $\mathrm{^{34}SO_2}$& &$1.8 \pm 0.3$&$18.0 \pm 0.7$& \\
  $\mathrm{^{33}SO_2}$& &0.2&4.0& \\
  $\mathrm{OS^{17}O}$& &$-$&$-$& \\
  $\mathrm{OS^{18}O}$& &$-$&$-$& \\
  $\mathrm{^{34}SO}$& &$2 \pm 1$&$18.1 \pm 0.5$&\multirow{3}{*}{$7.3 \pm 0.2$}\\
  $\mathrm{^{33}SO}$& &$0.32 \pm 0.01$&$16.4 \pm 0.1$& \\
  OCS& &$5 \pm 4$&$7.7 \pm 0.3$&\multirow{3}{*}{$7.7 \pm 0.2$}\\
  $\mathrm{O^{13}CS}$& &$0.14 \pm 0.03$&$6 \pm 5$& \\
  $\mathrm{^{13}CS}$& &$0.095 \pm 0.001$&$4.02 \pm 0.01$&$8.60 \pm 0.01$\\
  $\mathrm{H_2S}$& &$9.35$&$20.20$&$8.47$\\
  $\mathrm{H_2CS}$& &$0.4 \pm 0.1$&$4.6 \pm 0.4$&\multirow{2}{*}{$8.5 \pm 0.1$}\\
  $\mathrm{H_2C^{34}S}$& &$0.0057 \pm 0.0006$&$1 \pm 1$& \\
  $\mathrm{C^{18}O}$& &$23 \pm 6$&$6 \pm 2$&$8.8 \pm 0.2$\\
  \hline
  \multicolumn{5}{c}{mm3}\\
  \hline
  $\mathrm{SO_2}$&  &$26 \pm 8$&$18.3 \pm 0.5$&\multirow{5}{*}{$8.8 \pm 0.3$}\\
  $\mathrm{^{34}SO_2}$& &$1.3 \pm 0.2$&$18.3 \pm 0.5$& \\
  $\mathrm{^{33}SO_2}$& &$-$&$-$& \\
  $\mathrm{OS^{17}O}$& &$-$&$-$& \\
  $\mathrm{OS^{18}O}$& &$-$&$-$& \\
  $\mathrm{^{34}SO}$& &$2 \pm 1$&$16.8 \pm 0.5$&\multirow{3}{*}{$9.8 \pm 0.2$}\\
  $\mathrm{^{33}SO}$& &$0.31 \pm 0.01$&$18 \pm 17$& \\
  OCS& $107 \pm 9$ &$4 \pm 2$&$6.1 \pm 0.3$&\multirow{3}{*}{$8.6 \pm 0.2$}\\
  $\mathrm{O^{13}CS}$& &$0.23 \pm 0.06$&$6 \pm 2$& \\
  $\mathrm{^{13}CS}$& &$0.065 \pm 0.003$&$4.11 \pm 0.02$&$8.75 \pm 0.01$\\
  $\mathrm{H_2S}$& &$8.00$&$20.45$&$8.87$\\
  $\mathrm{H_2CS}$& &$0.50 \pm 0.01$&$4.09 \pm 0.01$&\multirow{2}{*}{$8.89 \pm 0.01$}\\
  $\mathrm{H_2C^{34}S}$& &$0.009 \pm 0.001$&$2.00$& \\
  $\mathrm{C^{18}O}$& &$17 \pm 5$&$6.1 \pm 0.7$&$8.5 \pm 0.2$\\
  \hline
  \multicolumn{5}{c}{mm4}\\
  \hline
  $\mathrm{SO_2}$& \multirow{14}{*}{$105 \pm 9$} &$24 \pm 12$&$14.4 \pm 0.7$&\multirow{5}{*}{$10.7 \pm 0.3$}\\
  $\mathrm{^{34}SO_2}$& &$1.1 \pm 0.3$&$14.9 \pm 0.6$& \\
  $\mathrm{^{33}SO_2}$& &$-$&$-$& \\
  $\mathrm{OS^{17}O}$& &$-$&$-$& \\
  $\mathrm{OS^{18}O}$& &$-$&$-$& \\
  $\mathrm{^{34}SO}$& &$1.5 \pm 0.7$&$12.1 \pm 0.5$&\multirow{3}{*}{$10.4 \pm 0.2$}\\
  $\mathrm{^{33}SO}$& &$0.23 \pm 0.01$&$12.8 \pm 1.6$& \\
  OCS&  &$4 \pm 3$&$6.5 \pm 0.4$&\multirow{3}{*}{$8.3 \pm 0.2$}\\
  $\mathrm{O^{13}CS}$& &$0.12 \pm 0.03$&$6 \pm 3$& \\
  $\mathrm{^{13}CS}$& &$0.087 \pm 0.001$&$4.36 \pm 0.03$&$8.48 \pm 0.01$\\
  $\mathrm{H_2S}$& &$5.60$&$16.27$&$10.44$\\
  $\mathrm{H_2CS}$& &$0.6 \pm 0.2$&$5.0 \pm 0.2$&\multirow{2}{*}{$8.3 \pm 0.1$}\\
  $\mathrm{H_2C^{34}S}$& &$0.008 \pm 0.001$&$2 \pm 1$& \\
  $\mathrm{C^{18}O}$& &$19 \pm 6$&$5.1 \pm 0.3$&$9.1 \pm 0.2$\\
  \hline
  \multicolumn{5}{c}{mm5}\\
  \hline
  $\mathrm{SO_2}$& \multirow{14}{*}{$105 \pm 11$} &$19 \pm 8$&$25.0 \pm 0.6$&\multirow{5}{*}{$10.8 \pm 0.3$}\\
  $\mathrm{^{34}SO_2}$& &$1.15 \pm 0.01$&$21.2 \pm 0.2$& \\
  $\mathrm{^{33}SO_2}$& &$-$&$-$& \\
  $\mathrm{OS^{17}O}$& &$-$&$-$& \\
  $\mathrm{OS^{18}O}$& &$-$&$-$& \\
  $\mathrm{^{34}SO}$& &$1.4 \pm 0.6$&$21.7 \pm 0.4$&\multirow{3}{*}{$11.2 \pm 0.2$}\\
  $\mathrm{^{33}SO}$& &$0.25 \pm 0.01$&$20 \pm 1$& \\
  OCS& &$3 \pm 2$&$7.9 \pm 0.4$&\multirow{3}{*}{$9.0 \pm 0.2$}\\
  $\mathrm{O^{13}CS}$& &$0.075 \pm 0.002$&$5.4 \pm 0.7$& \\
  $\mathrm{^{13}CS}$& &$0.051 \pm 0.001$&$6.3 \pm 0.5$&$8.94 \pm 0.02$\\
  $\mathrm{H_2S}$& &$6.81$&$26.13$&$10.05$\\
  $\mathrm{H_2CS}$& &$0.34 \pm 0.01$&$4.70 \pm 0.03$&\multirow{2}{*}{$8.82 \pm 0.01$}\\
  $\mathrm{H_2C^{34}S}$& &$0.006 \pm 0.001$&$2.00$& \\
  $\mathrm{C^{18}O}$& &$12 \pm 3$&$6.3 \pm 0.5$&$9.1 \pm 0.2$\\
  \hline
  \multicolumn{5}{c}{mm7}\\
  \hline
  $\mathrm{SO_2}$&  &$0.8 \pm 0.1$&$7.9 \pm 0.6$&\multirow{5}{*}{$10.7 \pm 0.4$}\\
  $\mathrm{^{34}SO_2}$& &$0.022 \pm 0.006$&$15 \pm 5$& \\
  $\mathrm{^{33}SO_2}$& &$-$&$-$& \\
  $\mathrm{OS^{17}O}$& &$-$&$-$& \\
  $\mathrm{OS^{18}O}$& &$-$&$-$& \\
  $\mathrm{^{34}SO}$& &$0.033 \pm 0.001$&$6 \pm 1$&\multirow{3}{*}{$12 \pm 12$}\\
  $\mathrm{^{33}SO}$& &$-$&$-$& \\
  OCS&$68 \pm 11$ &$0.8\pm 0.3$&$2.8 \pm 0.4$&\multirow{3}{*}{$8.9 \pm 0.2$}\\
  $\mathrm{O^{13}CS}$& &$0.010 \pm 0.001$&$2 \pm 2$& \\
  $\mathrm{^{13}CS}$& &$0.015 \pm 0.001$&$3.2 \pm 0.2$&$8.78 \pm 0.02$\\
  $\mathrm{H_2S}$& &$0.085 \pm 0.004$&$2.6 \pm 0.4$&$9.2 \pm 0.3$\\
  $\mathrm{H_2CS}$& &$0.12 \pm 0.01$&$2.62 \pm 0.03$&\multirow{2}{*}{$9.05 \pm 0.02$}\\
  $\mathrm{H_2C^{34}S}$& &$-$&$-$& \\
  $\mathrm{C^{18}O}$& &$5.66 \pm 0.07$&$3 \pm 3$&$8.99 \pm 0.02$\\
\enddata
\end{deluxetable}

\begin{deluxetable*}{crrrrrrr}
\tablecaption{Sulfur isotopologue ratios towards different region.\label{tab:sulfur ratio}}
\tablewidth{700pt}
\tabletypesize{\scriptsize}
\tablehead{\colhead{Ratio} & \colhead{Hot core} & \colhead{mm2} & \colhead{mm3} & \colhead{mm4} & \colhead{mm5} & \colhead{mm7} & \colhead{average} }
\startdata
  $\mathrm{SO_2 /^{34}SO_2}$&$10 \pm 3$&$17 \pm 8$&$20 \pm 7$&$22 \pm 12$&$17 \pm 7$&$35 \pm 12$&$20 \pm 4$\\
  $\mathrm{^{34}SO/^{33}SO}$&$3 \pm 2$&$6 \pm 3$&$6 \pm 3$&$6 \pm 3$&$6 \pm 2$&$-$& $6 \pm 1$ \\
  $\mathrm{OCS/O^{13}CS}$&$18 \pm 14$&$38 \pm 30$&$18 \pm 10$&$36 \pm 26$&$43 \pm 21$&$75 \pm 28$& $38 \pm 9$\\
  $\mathrm{H_2CS/H_2C^{34}S}$&$19 \pm 8$&$67 \pm 21$&$56 \pm 6$&$69 \pm 27$&$57 \pm 10$&$-$& $53 \pm 7$\\
\enddata
\end{deluxetable*}

\begin{deluxetable*}{crrrrrr}
\tablecaption{Abundance toward different region.\label{tab:abundance}}
\tablewidth{700pt}
\tabletypesize{\scriptsize}
\tablehead{\colhead{Species} & \colhead{Hot core} & \colhead{mm2} & \colhead{mm3} & \colhead{mm4} & \colhead{mm5} & \colhead{mm7} }
\startdata
  $\mathrm{SO_2}$&$3\pm2 \times 10^{-7}$&$3\pm1 \times 10^{-7}$&$2.4\pm0.9 \times 10^{-7}$&$2\pm1 \times 10^{-7}$&$2\pm1 \times 10^{-7}$&$2.7\pm0.5 \times 10^{-8}$\\
  $\mathrm{^{34}SO_2}$&$3\pm2 \times 10^{-8}$&$1.6\pm0.5 \times 10^{-8}$&$1.2\pm0.3 \times 10^{-8}$&$10\pm4 \times 10^{-9}$&$1.4\pm0.3 \times 10^{-8}$&$8\pm2 \times 10^{-10}$\\
  $\mathrm{^{33}SO_2}$&$1.0\pm0.5 \times 10^{-9}$&$1.8\pm0.4 \times 10^{-9}$&$-$&$-$&$-$&$-$\\
  $\mathrm{OS^{17}O}$&$4\pm2 \times 10^{-10}$&$-$&$-$&$-$&$-$&$-$\\
  $\mathrm{OS^{18}O}$&$1.0\pm0.5 \times 10^{-9}$&$-$&$-$&$-$&$-$&$-$\\
  $\mathrm{^{34}SO}$&$2\pm1 \times 10^{-8}$&$2\pm1 \times 10^{-8}$&$2\pm1 \times 10^{-8}$&$1.3\pm0.8 \times 10^{-8}$&$1.8\pm0.8 \times 10^{-8}$&$1.17\pm0.04 \times 10^{-9}$\\
  $\mathrm{^{33}SO}$&$7\pm4 \times 10^{-9}$&$2.8\pm0.7 \times 10^{-9}$&$2.8\pm0.7 \times 10^{-9}$&$2.1\pm0.6 \times 10^{-9}$&$3.1\pm0.7 \times 10^{-9}$&$-$\\
  $\mathrm{OCS}$&$5\pm4 \times 10^{-8}$&$5\pm4 \times 10^{-8}$&$4\pm2 \times 10^{-8}$&$4\pm3 \times 10^{-8}$&$4\pm2 \times 10^{-8}$&$3\pm1 \times 10^{-8}$\\
  $\mathrm{O^{13}CS}$&$3\pm2 \times 10^{-9}$&$1.2\pm0.4 \times 10^{-9}$&$2\pm2 \times 10^{-9}$&$1.1\pm0.4 \times 10^{-9}$&$9\pm2 \times 10^{-10}$&$3.5\pm0.4 \times 10^{-10}$\\
  $\mathrm{^{13}CS}$&$1\pm1 \times 10^{-9}$&$8\pm2 \times 10^{-10}$&$6\pm1 \times 10^{-10}$&$8\pm2 \times 10^{-10}$&$6\pm1 \times 10^{-10}$&$5.3\pm0.4 \times 10^{-10}$\\
  $\mathrm{H_2S}$&$9\pm6 \times 10^{-8}$&$8\pm2
   \times 10^{-8}$&$7\pm2 \times 10^{-8}$&$5\pm1 \times 10^{-8}$&$9\pm2 \times 10^{-8}$&$3.0\pm0.2 \times 10^{-9}$\\
  $\mathrm{H_2CS}$&$4\pm3 \times 10^{-9}$&$3\pm1 \times 10^{-9}$&$5\pm1 \times 10^{-9}$&$5\pm2 \times 10^{-9}$&$4.3\pm0.9 \times 10^{-9}$&$4.2\pm0.4 \times 10^{-9}$\\
  $\mathrm{H_2C^{34}S}$&$2\pm1 \times 10^{-10}$&$5\pm1 \times 10^{-11}$&$8\pm2 \times 10^{-10}$&$8\pm2 \times 10^{-11}$&$8\pm2 \times 10^{-10}$&$-$\\
\enddata
\end{deluxetable*}


\begin{figure*}
\newcounter{1}
\setcounter{1}{\value{figure}}
\setcounter{figure}{0}
\renewcommand\thefigure{A.\arabic{figure}}
  \centering
  \includegraphics[width=1.0\linewidth,trim=0 0 0 0,clip]{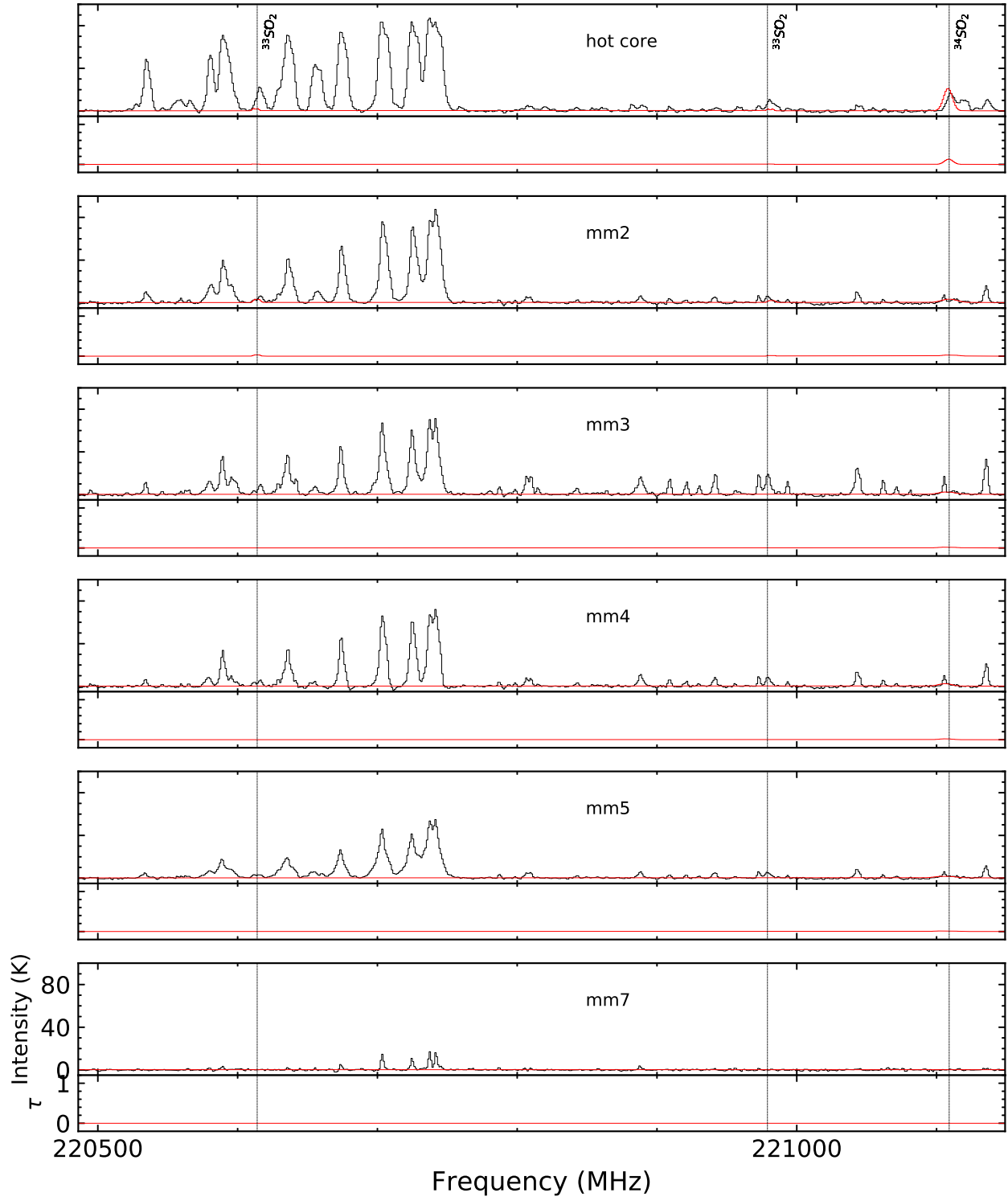}
        \caption{Spectral surveys for different regions. The six subfigures show the spectra and optical depth of the identified molecules for six positions. For each subfigure, the black curve in the upper panel shows the combined data, and the red curve shows the model spectra obtained with XCLASS. The red curve in the lower panel shows the optical depth of each line.}
         \label{fig:lines}
\setcounter{figure}{\value{1}}
\end{figure*}

\begin{figure*}
\setcounter{1}{\value{figure}}
\setcounter{figure}{0}
\renewcommand\thefigure{A.\arabic{figure}}
\begin{center}

\includegraphics[width=1.0\linewidth,trim=0 0 0 0,clip]{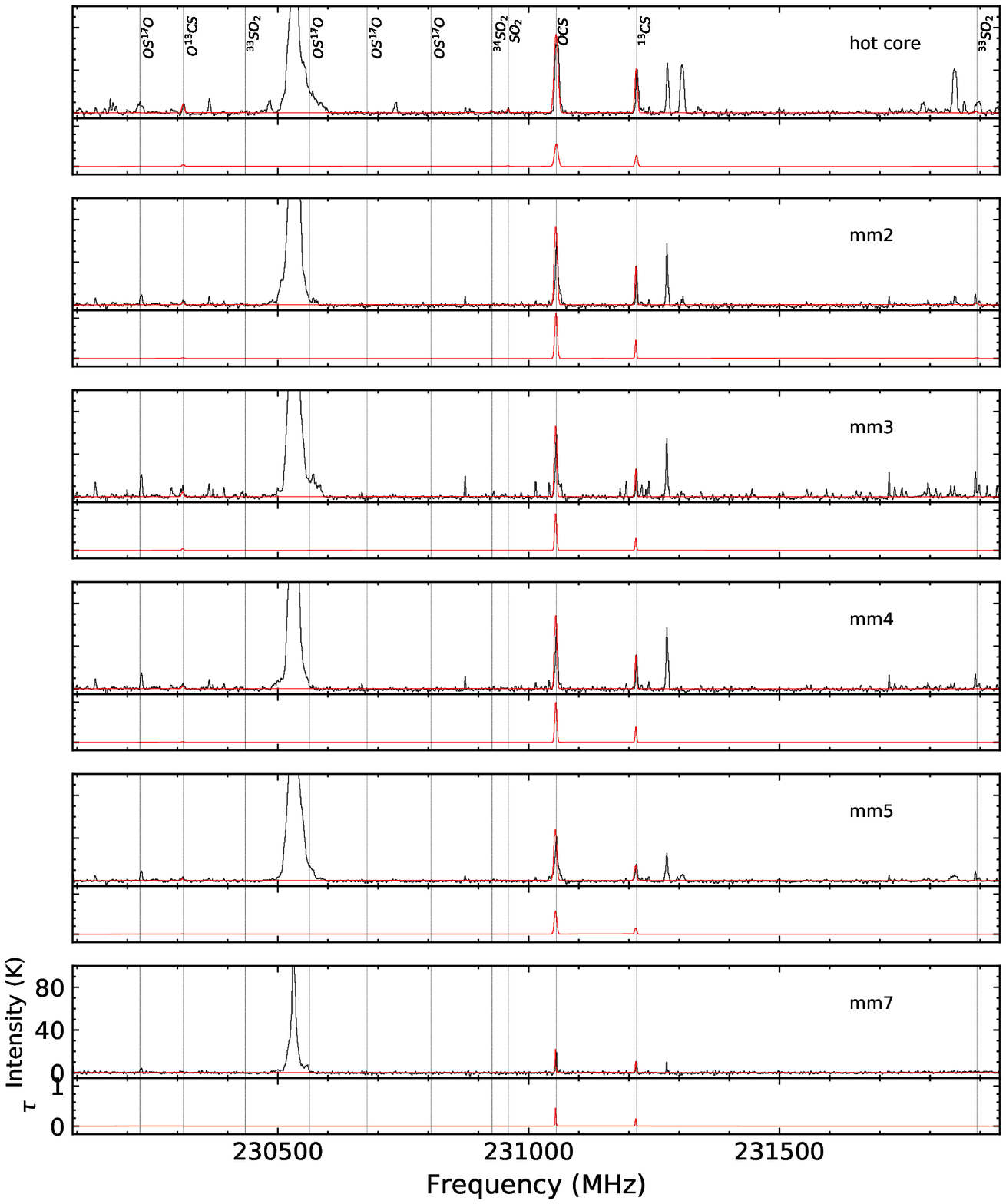}

\caption{(continued)}
\end{center}
\setcounter{figure}{\value{1}}
\end{figure*}

\begin{figure*}
\setcounter{1}{\value{figure}}
\setcounter{figure}{0}
\renewcommand\thefigure{A.\arabic{figure}}
\begin{center}

\includegraphics[width=1.0\linewidth,trim=0 0 0 0,clip]{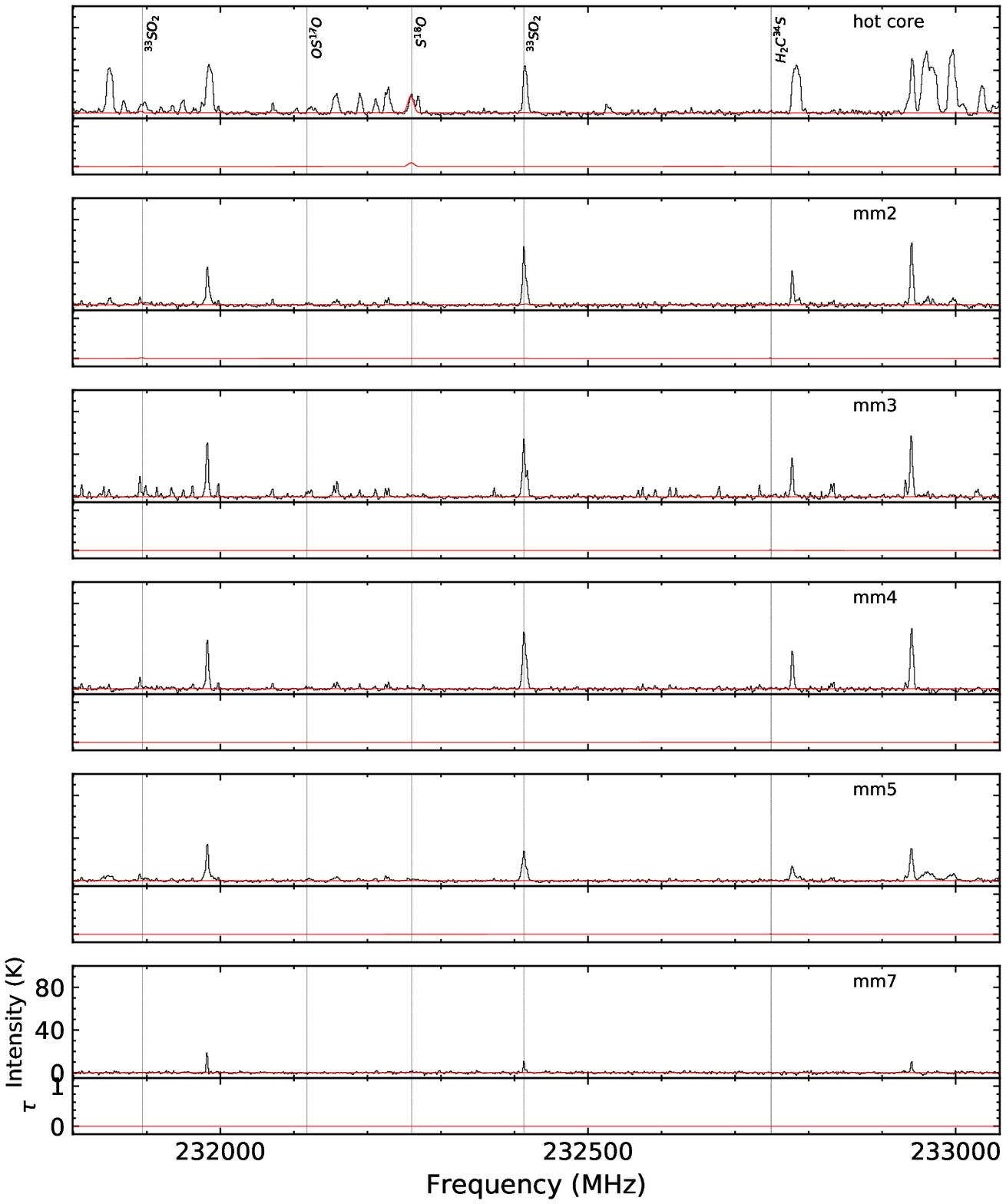}

\caption{(continued)}
\end{center}
\setcounter{figure}{\value{1}}
\end{figure*}

\begin{figure*}
\setcounter{1}{\value{figure}}
\setcounter{figure}{0}
\renewcommand\thefigure{A.\arabic{figure}}
\begin{center}

\includegraphics[width=1.0\linewidth,trim=0 0 0 0,clip]{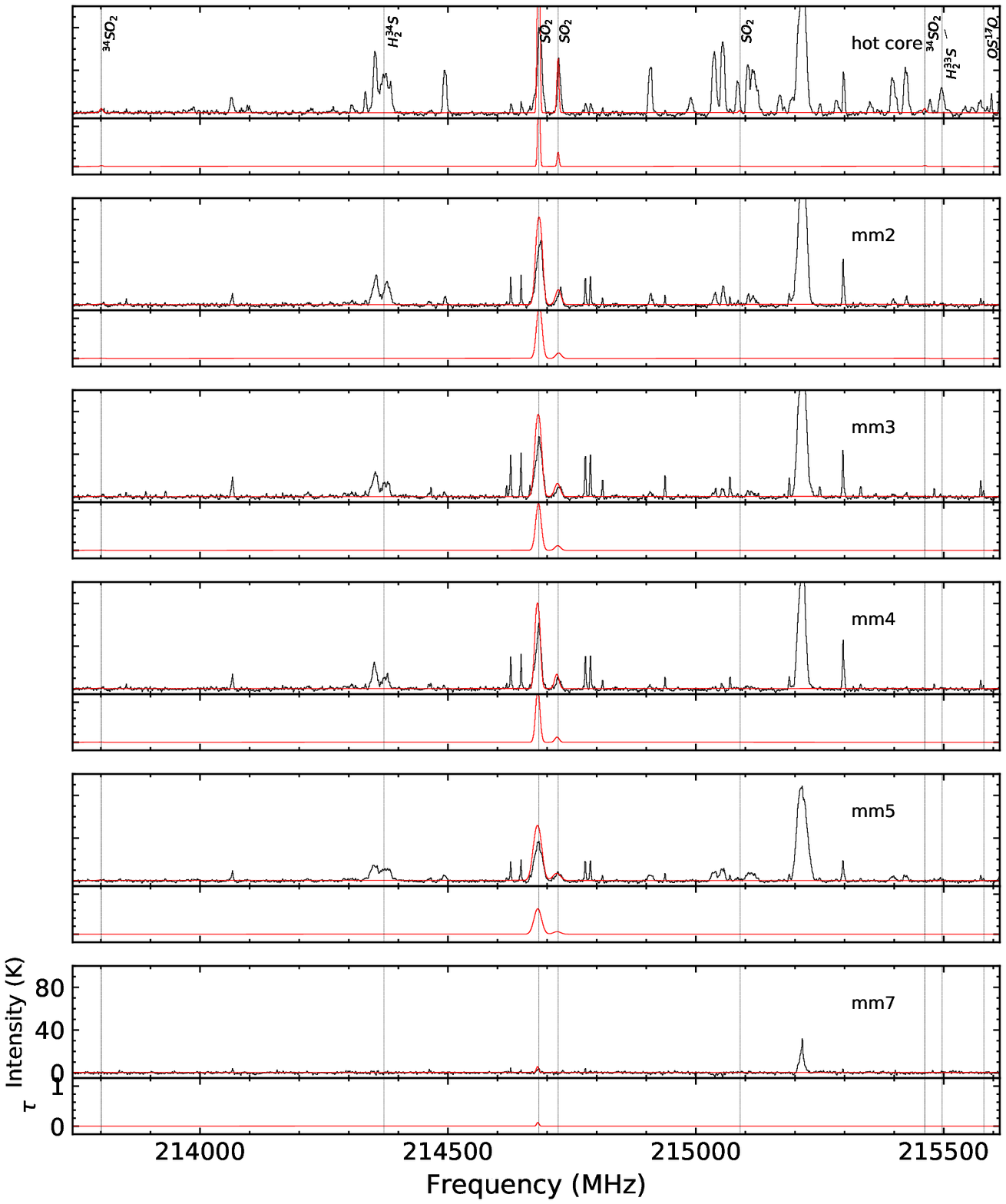}

\caption{(continued)}
\end{center}
\setcounter{figure}{\value{1}}
\end{figure*}

\begin{figure*}
\setcounter{1}{\value{figure}}
\setcounter{figure}{0}
\renewcommand\thefigure{A.\arabic{figure}}
\begin{center}

\includegraphics[width=1.0\linewidth,trim=0 0 0 0,clip]{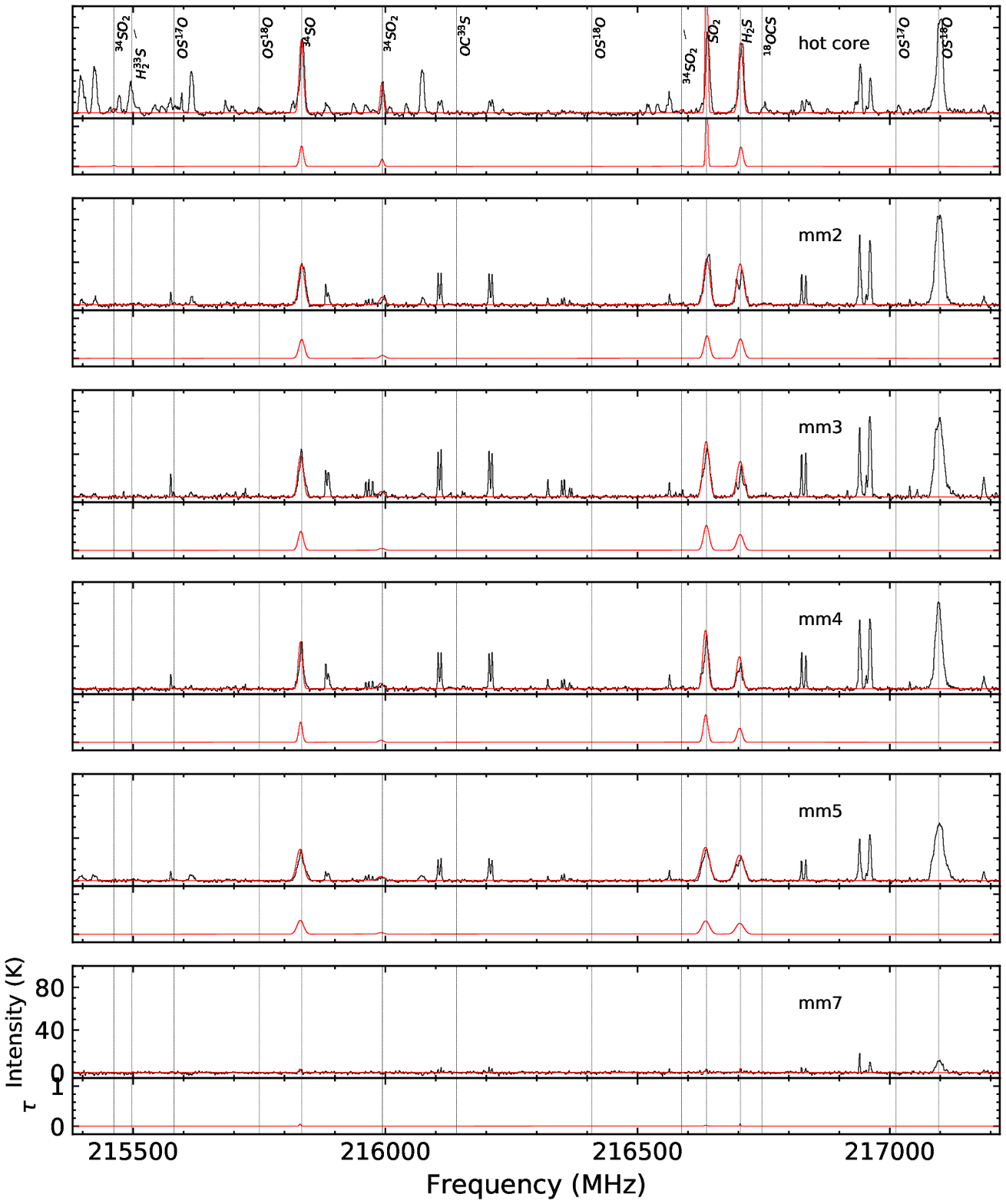}

\caption{(continued)}
\end{center}
\setcounter{figure}{\value{1}}
\end{figure*}

\begin{figure*}
\setcounter{1}{\value{figure}}
\setcounter{figure}{0}
\renewcommand\thefigure{A.\arabic{figure}}
\begin{center}

\includegraphics[width=1.0\linewidth,trim=0 0 0 0,clip]{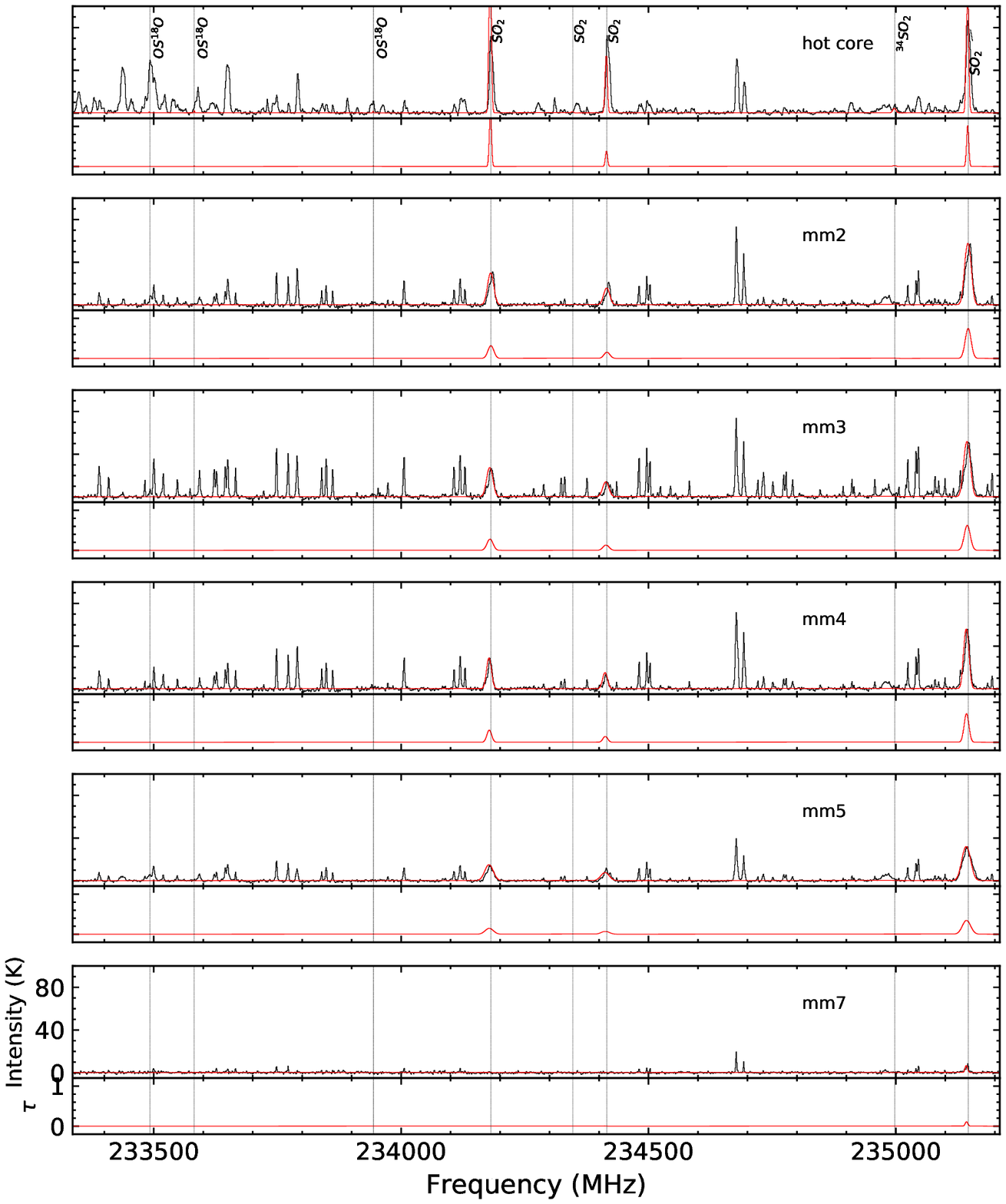}

\caption{(continued)}
\end{center}
\setcounter{figure}{\value{1}}
\end{figure*}

\begin{figure*}
\setcounter{1}{\value{figure}}
\setcounter{figure}{0}
\renewcommand\thefigure{A.\arabic{figure}}
\begin{center}

\includegraphics[width=1.0\linewidth,trim=0 0 0 0,clip]{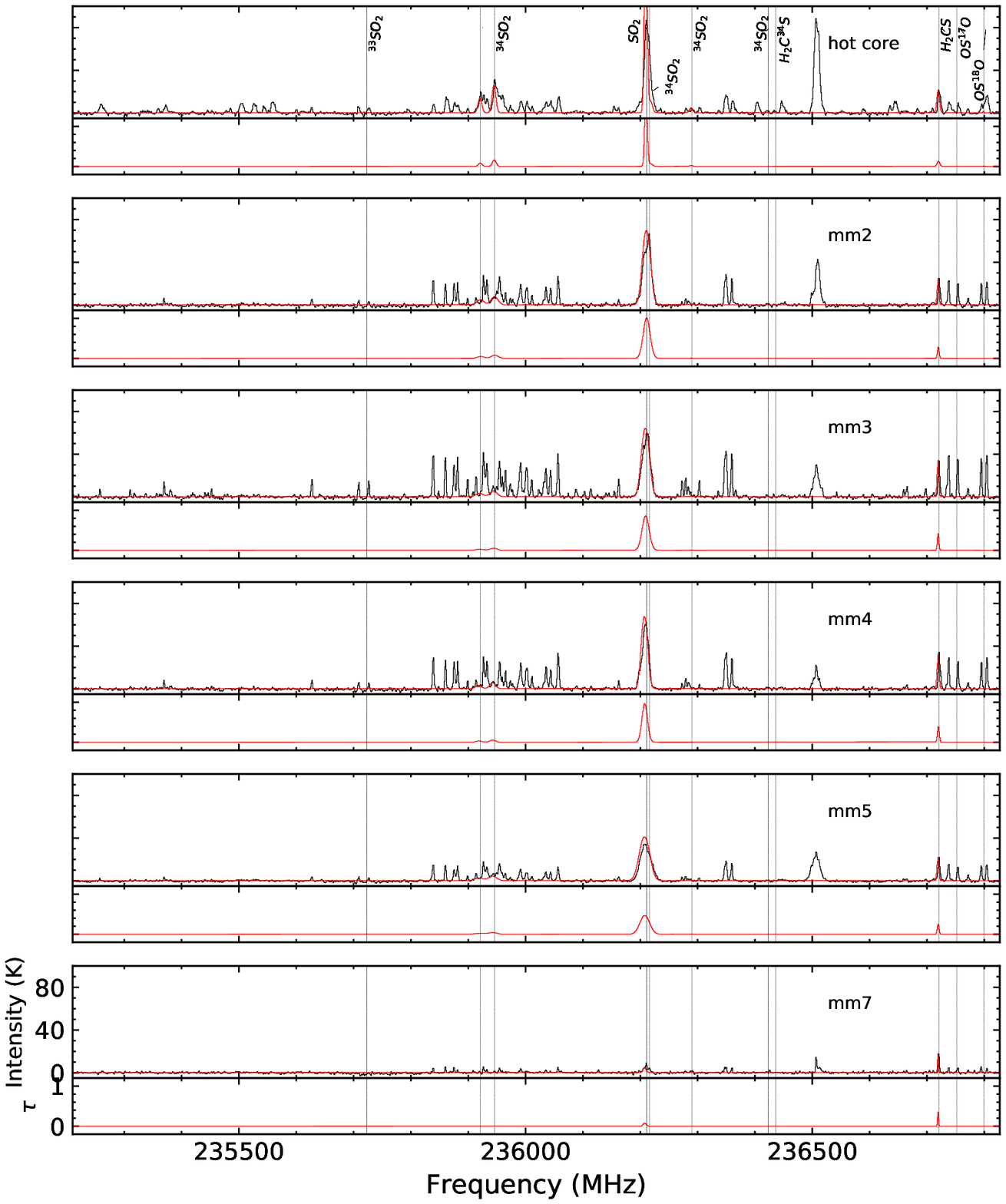}

\caption{(continued)}
\end{center}
\setcounter{figure}{\value{1}}
\end{figure*}

\begin{figure*}
\setcounter{1}{\value{figure}}
\setcounter{figure}{0}
\renewcommand\thefigure{A.\arabic{figure}}
\begin{center}

\includegraphics[width=1.0\linewidth,trim=0 0 0 0,clip]{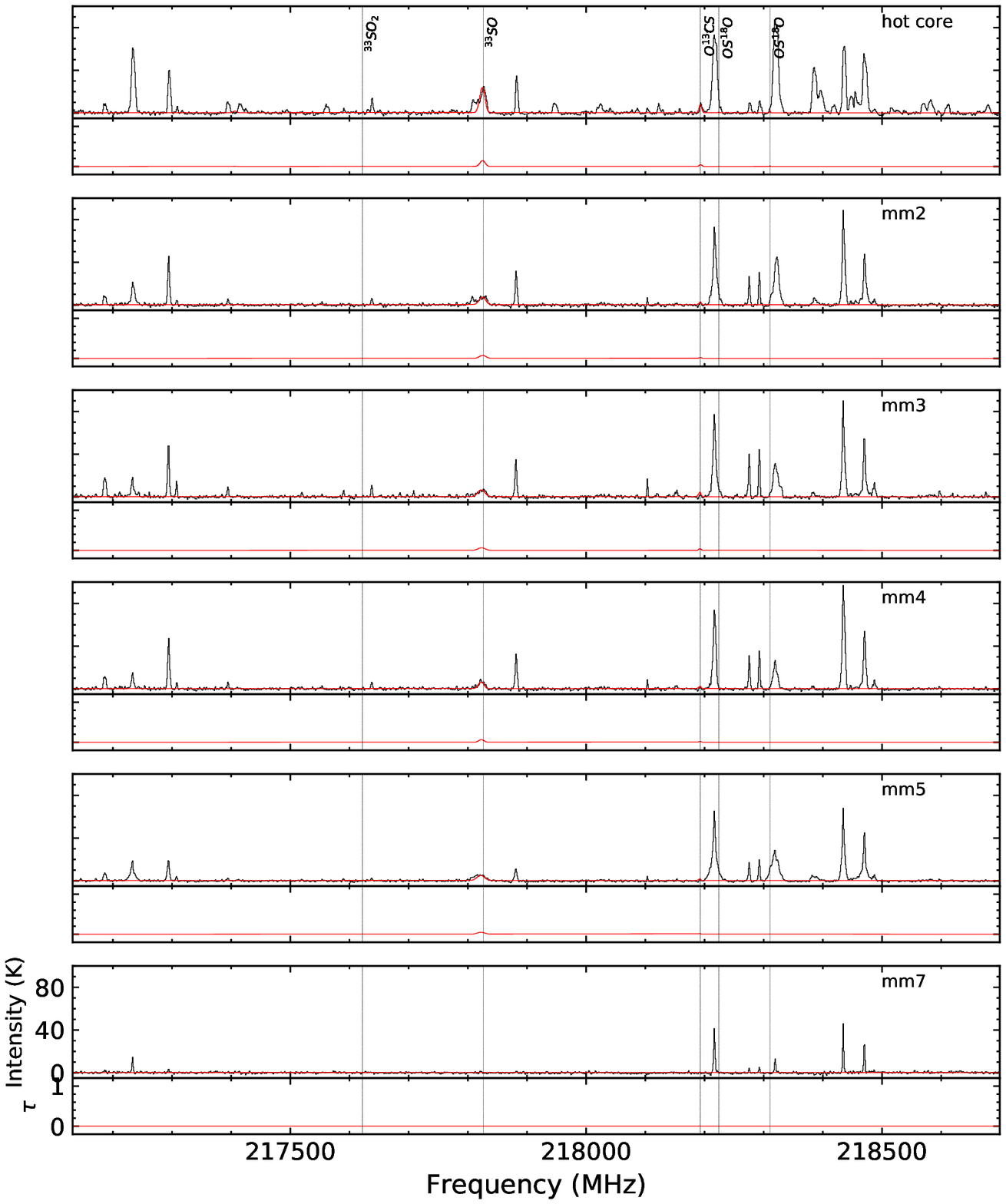}

\caption{(continued)}
\end{center}
\setcounter{figure}{\value{1}}
\end{figure*}

\begin{figure*}
\setcounter{1}{\value{figure}}
\setcounter{figure}{0}
\renewcommand\thefigure{A.\arabic{figure}}
\begin{center}

\includegraphics[width=1.0\linewidth,trim=0 0 0 0,clip]{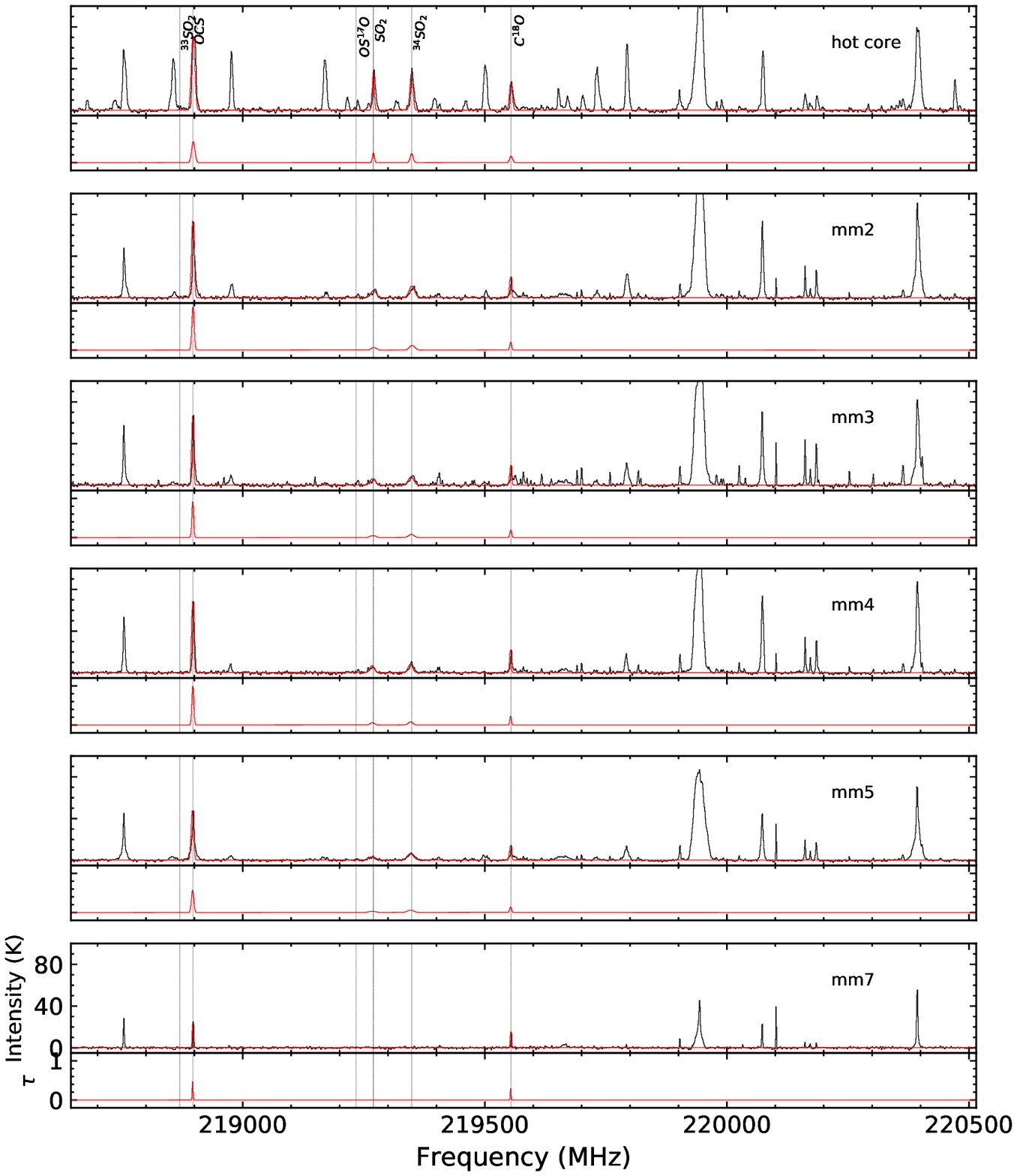}

\caption{(continued)}
\end{center}
\setcounter{figure}{\value{1}}
\end{figure*}

\begin{figure*}
\setcounter{1}{\value{figure}}
\setcounter{figure}{0}
\renewcommand\thefigure{A.\arabic{figure}}
\begin{center}

\includegraphics[width=1.0\linewidth,trim=0 0 0 0,clip]{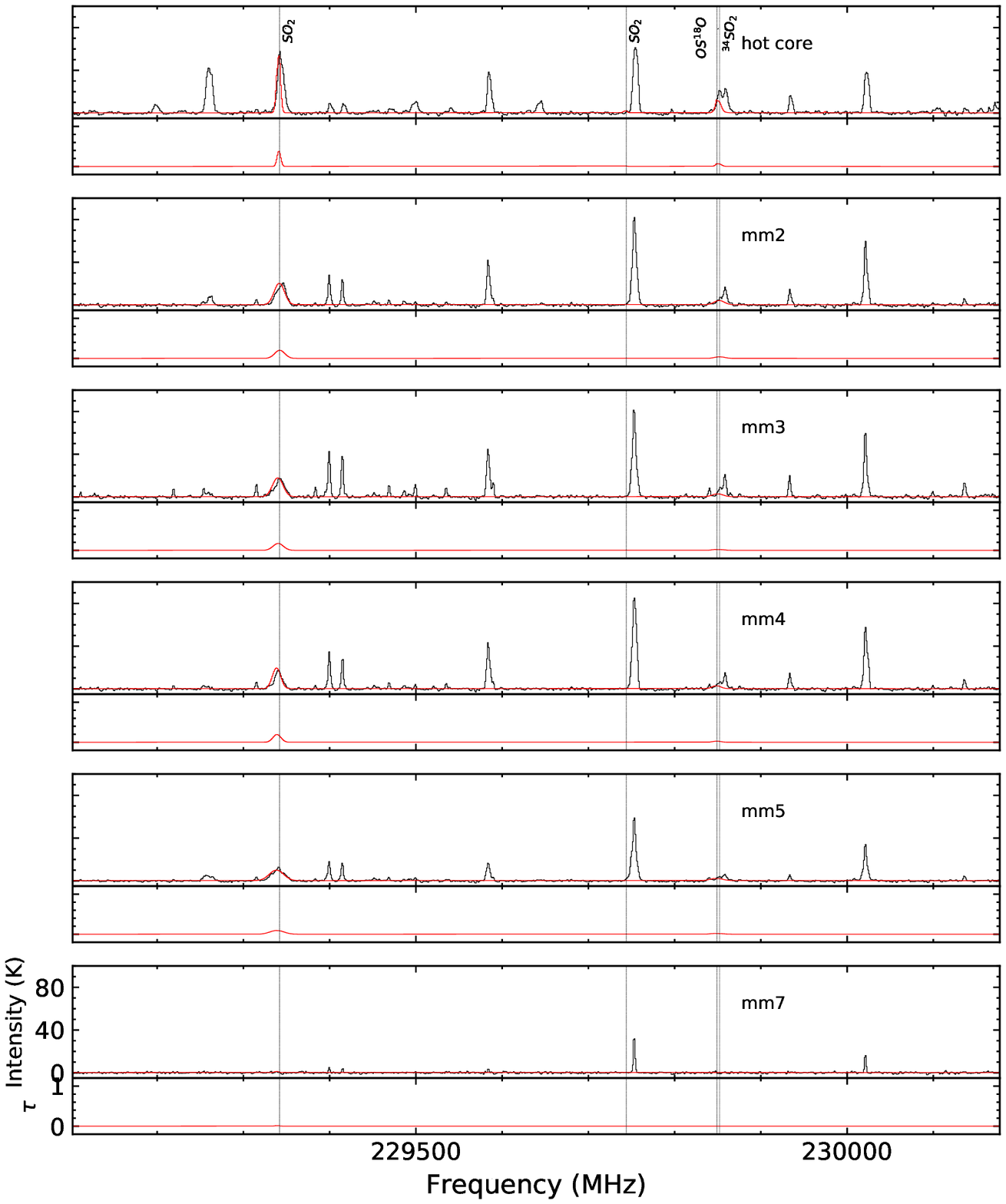}

\caption{(continued)}
\end{center}
\setcounter{figure}{\value{1}}
\end{figure*}

\begin{figure*}
\newcounter{2}
\setcounter{2}{\value{figure}}
\setcounter{figure}{1}
\renewcommand\thefigure{A.\arabic{figure}}
  \centering
  \includegraphics[width=4.0in,trim=0 0 0 0,clip,angle=-90]{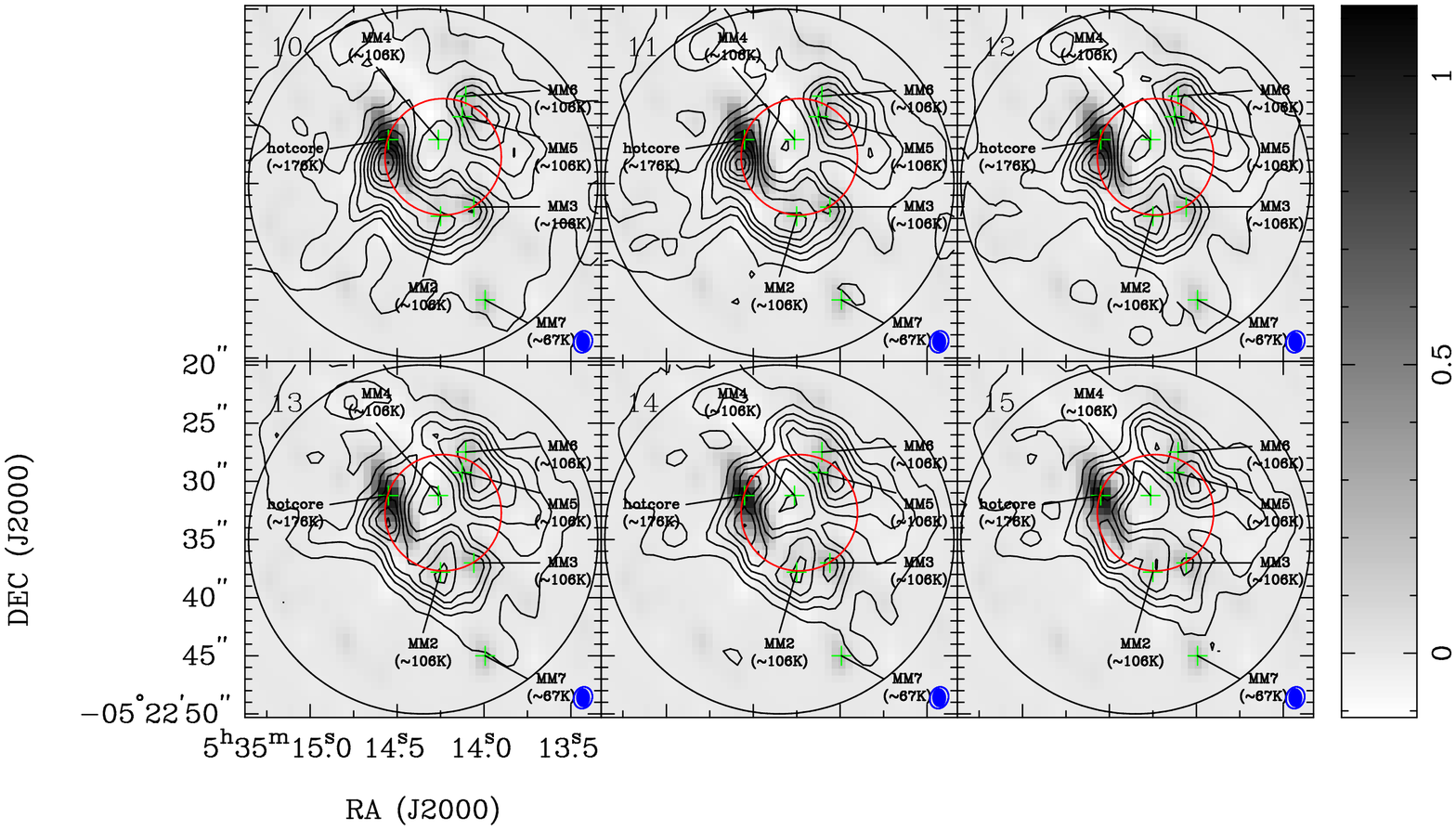}
        \caption{Channel maps of $\mathrm{SO_2}$, over a velocity range of 10 to 15 \kms. The gray map in the background shows the ALMA-only continuum emission. The peak positions of the continuum emission are marked with green crosses. Black contours show the line emission from the ALMA-30\,m combination, starting from 5\% of the emission peak and increasing by steps of 5\%. The red circle indicates the ring-like structure.\label{fig:SO2channelmap}}
\setcounter{figure}{\value{2}}
\end{figure*}

\begin{figure*}
\newcounter{3}
\setcounter{3}{\value{figure}}
\setcounter{figure}{2}
\renewcommand\thefigure{A.\arabic{figure}}
  \centering
  \includegraphics[width=4.0in,trim=0 0 0 0,clip,angle=-90]{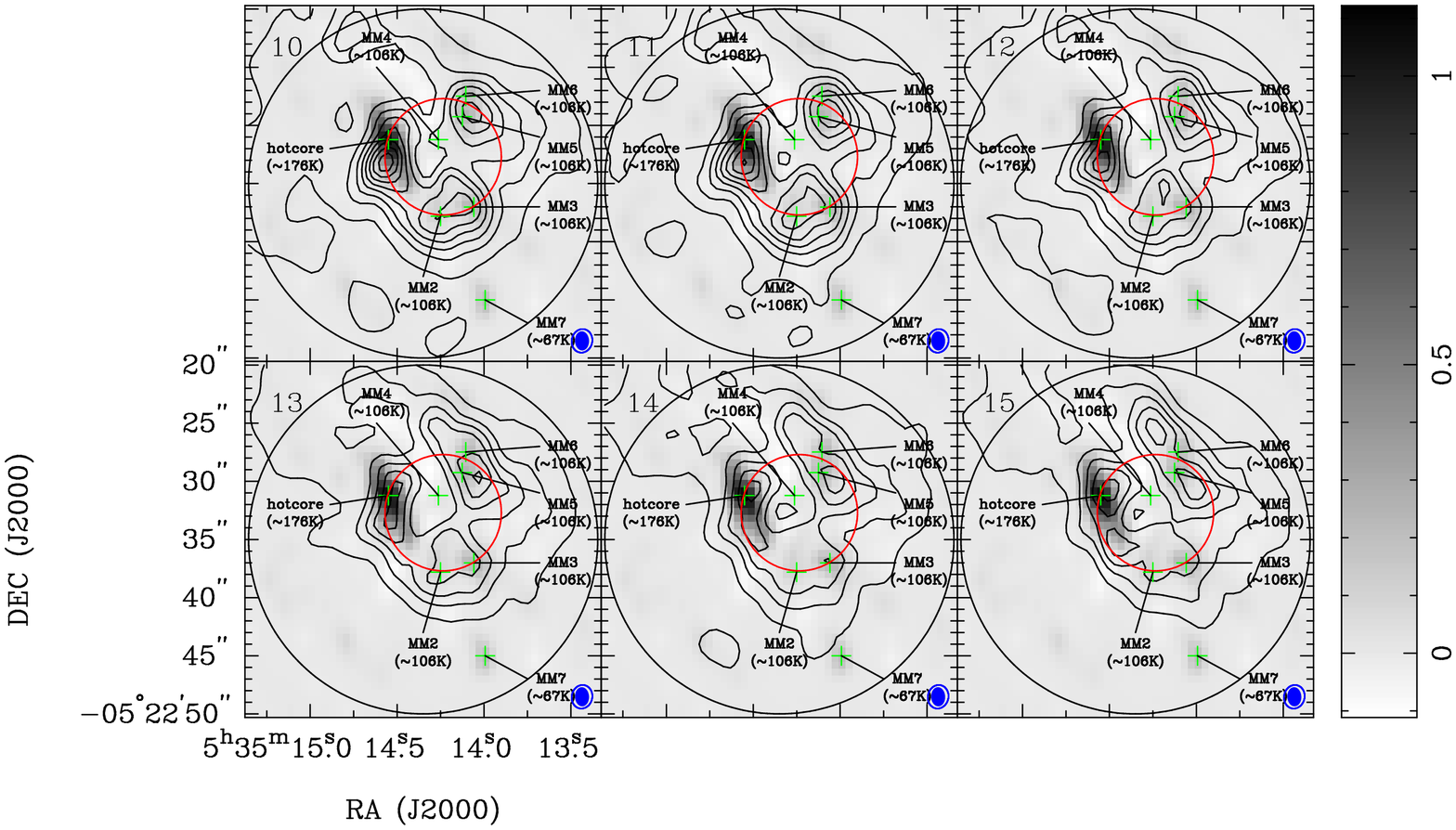}
        \caption{Channel maps of $\mathrm{^{34}SO}$, over a velocity range of 10 to 15 \kms. The gray map in the background shows the ALMA-only continuum emission. The peak positions of the continuum emission are marked with green crosses. Black contours show the line emission from the ALMA-30\,m combination, starting from 5\% of the emission peak and increasing by steps of 5\%. The red circle indicates the ring-like structure.\label{fig:34SOchannelmap}}
\setcounter{figure}{\value{3}}
\end{figure*}

\begin{figure*}
\newcounter{4}
\setcounter{4}{\value{figure}}
\setcounter{figure}{3}
\renewcommand\thefigure{A.\arabic{figure}}
  \centering
  \includegraphics[width=1.0\linewidth]{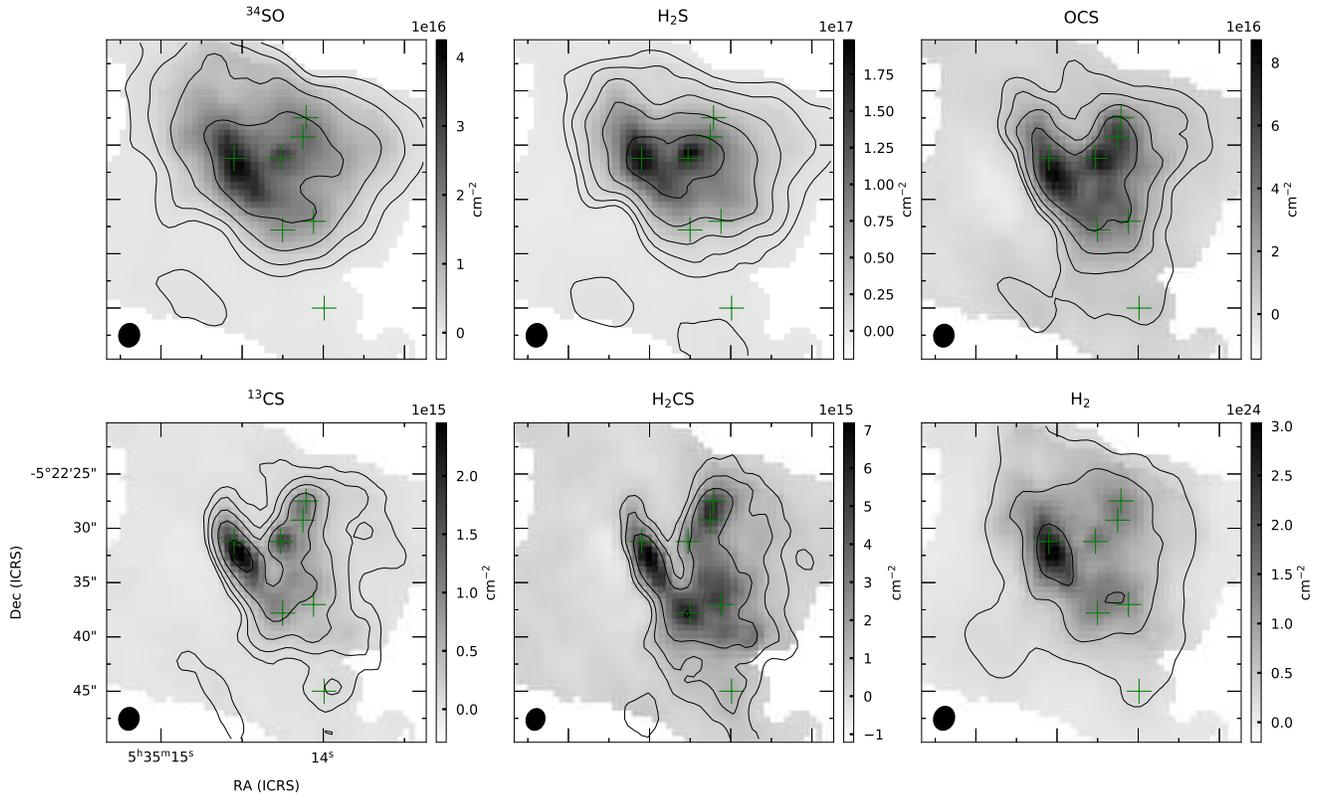}
        \caption{The column density distributions of $^{34}$SO, H$_2$S, OCS, $^{13}$CS, H$_2$CS, and H$_2$. The black contours show the integrated intensity of each molecule in Figure \ref{fig:gasmap}. The contour levels are 1, 2, 4, 8, 16 $\times$ 5$\sigma$. Green crosses indicate the continuum peaks. Regions outside the pixel values corresponding to a 3$\sigma$ detection of the most extended integrated intensity of SO$_2$ lines are masked out. \label{fig:col_map}}
\setcounter{figure}{\value{4}}
\end{figure*}

\end{document}